\documentclass[runningheads]{llncs}
\usepackage{times}
\usepackage{microtype}
\usepackage{comment}
\usepackage{adjustbox}
\usepackage{listings}
\usepackage{lstautogobble}
\usepackage{booktabs}
\usepackage{makecell}
\usepackage{multirow}
\usepackage{pdflscape}
\usepackage{epstopdf}
\usepackage[font={small,normalfont},
labelfont=normalfont,
format=hang,
format=plain,
margin=0pt,
width=0.8\textwidth,
]{caption}
\usepackage[list=true, labelformat=simple]{subcaption}

\usepackage{color}
\usepackage{zi4}
\usepackage[utf8]{inputenc}
\usepackage[algo2e,ruled,lined, boxed]{algorithm2e}
\newcommand{\cmr}{\fontfamily{cmr}\selectfont}
\SetAlFnt{\cmr\scriptsize} \usepackage{ stmaryrd }
\usepackage{paralist}
\usepackage[inline]{enumitem}
\usepackage{amssymb}
\usepackage{amsmath}
\usepackage{url}
\usepackage{xspace}
\usepackage{paralist}
\usepackage{stmaryrd}
\usepackage{textcomp}
\usepackage{colortbl}
\PassOptionsToPackage{dvipsnames}{xcolor}

\newcommand{\ag}[1]{\textcolor{RoyalPurple}{[{\bf AG:} #1]}}

\newcommand{\sharon}[1]{[\textcolor{red}{\textbf{Sharon}: #1}]}
\newcommand{\jeff}[1]{[\textcolor{SeaGreen}{\textbf{JEF}: #1}]}

\renewcommand{\ag}[1]{}
\renewcommand{\jeff}[1]{}

\newcommand{\limp}{\Rightarrow}

\newcommand{\Sig}{\Sigma}

\newcommand{\lit}{\textit{lit}}

\newcommand{\Land}{\bigwedge}
\newcommand{\Lor}{\bigvee}

\newcommand{\smt}{\textsc{SMT}\xspace}

\newcommand{\safe}{\textsc{safe}\xspace}
\newcommand{\unsafe}{\textsc{unsafe}\xspace}

\newcommand{\true}{\mathit{true}}

\newcommand{\Tr}{\mathit{Tr}}
\newcommand{\Bad}{\mathit{Bad}}
\newcommand{\Init}{\mathit{Init}}
\newcommand{\Inv}{\mathit{Inv}}

\newcommand{\Ict}{\textsc{IC3}\xspace}

\newcommand{\ints}{\mathbb{Z}}

\newcommand{\cF}{\mathcal{F}}

\newcommand{\cC}{\mathcal{C}}

\newcommand{\cL}{\mathcal{L}}

\newcommand{\cS}{\mathcal{S}}
\newcommand{\cO}{\mathcal{O}}
\newcommand{\cU}{\mathcal{U}}
\newcommand{\cT}{\mathcal{T}}

\newcommand{\Spacer}{\textsc{Spacer}\xspace}
\newcommand{\GSpacer}{\textsc{GSpacer}\xspace}
\newcommand{\csm}{\textsc{GSpacer}\xspace}
\newcommand{\ddchc}{\textsc{LArb}\xspace}
\newcommand{\concr}{\textsc{Concretize}}
\newcommand{\concrlit}{\textsc{Concretize\_lit}}

\newcommand{\fVar}[1]{\textsc{Vars}(#1)}
\newcommand{\lind}[1]{#1^{L_{\downarrow}}}
\newcommand{\const}[1]{\textsc{Consts}(#1)}
\newcommand{\coeff}[2]{\textsc{coeff}(#1, #2)}
\newcommand{\pobclstr}[2]{\cC_{\textit{pob}}(\langle #1, #2\rangle)}
\newcommand{\lclstr}[1]{\cC_{\textit{lemma}}(#1)}

\newcommand{\conjProc}[3]{\textsc{Conjecture}(#1, #2, #3)}
\newcommand{\isSat}{\textsc{isSat}}
\newcommand{\nonLin}{\textsc{nonLin}}
\newcommand{\mbp}{\textsc{MBP}}
\newcommand{\itp}{\textsc{ITP}}
\newcommand{\gen}{\textsc{GEN}}

\newcommand{\pob}{\textsc{pob}\xspace}
\newcommand{\pobs}{\textsc{pob}s\xspace}

\newcommand{\Vars}{\vec{v}}
\newcommand{\Consts}{\vec{x}}
\newcommand{\Nums}{\vec{n}}
\newcommand{\Alphas}{\vec{\alpha}}
\newcommand{\nau}[2]{\textit{anti}(#1, #2)}
\newcommand{\coup}[3]{#1 \bowtie_{#3} #2}
\newcommand{\iso}[2]{\textsc{Iso}(#1, #2)}
\newcommand{\lia}{\textsc{LIA}\xspace}
\newcommand{\liamod}{\ensuremath{\textsc{LIA}^{-\text{div}}}\xspace}
\renewcommand{\gets}{:=}

\newcommand{\mc}{\multicolumn}

\DeclareMathOperator{\kernel}{kernel}
\DeclareMathOperator{\rank}{rank}
\SetKw{Assert}{assert}

\newcommand{\coord}[1]{#1^\text{th}}

\renewcommand{\paragraph}[1]{\vspace{0.02in}\noindent\emph{#1}}
 
\usepackage{mathtools}
\usepackage[hidelinks]{hyperref}
\usepackage{cleveref}
\usepackage{tikz}
\usetikzlibrary{shapes, fit, calc}
\title{Global Guidance for Local Generalization in Model Checking}
\author{\author{Hari Govind V K\inst{1} \and YuTing Chen\inst{2} \and Sharon Shoham\inst{3} \and Arie Gurfinkel\inst{1}}}
\institute{University of Waterloo \and Chalmers University of Technology \and Tel Aviv University}
\date{}
\usepackage[firstpage]{draftwatermark}\SetWatermarkText{\hspace*{5.5in}\raisebox{9in}{\includegraphics[scale=0.1]{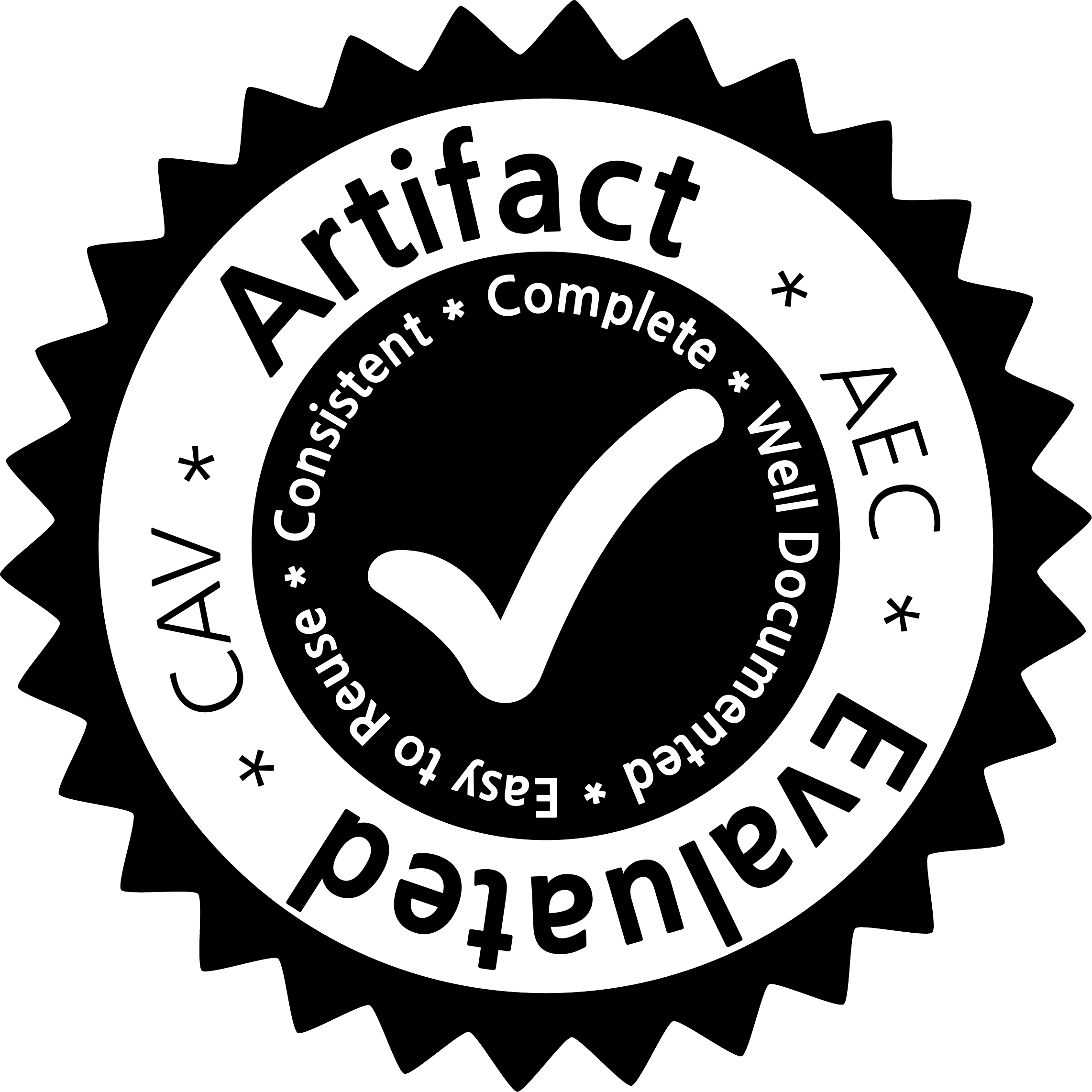}}}\SetWatermarkAngle{0} 

\Crefname{algorithm}{Alg.}{Alg.}
\crefname{algorithm}{Alg.}{Alg.}
\Crefname{section}{Sec.}{Sec.}
\crefname{section}{Sec.}{Sec.}
\Crefname{figure}{Fig.}{Fig.}
\crefname{figure}{Fig.}{Fig.}
\Crefname{equation}{Eq.}{Eq.}

\definecolor{bluekeywords}{rgb}{0.13, 0.13, 1}
\definecolor{greencomments}{rgb}{0, 0.5, 0}
\definecolor{redstrings}{rgb}{0.9, 0, 0}
\definecolor{graynumbers}{rgb}{0.5, 0.5, 0.5}
\begin{document}
\AlgoDontDisplayBlockMarkers
\SetAlgoNoEnd
\SetAlgoNoLine
\DontPrintSemicolon
\SetInd{0.2em}{0.2em}
\SetKwProg{Fn}{function}{:}{}
\SetKwInput{KwIn}{In}
\SetKwInput{KwOut}{Out}

\maketitle
\begin{abstract}
  \smt-based model checkers, especially \Ict-style ones, are currently the most
  effective techniques for verification of infinite state systems. They infer
  \emph{global} inductive invariants via \emph{local} reasoning about a single
  step of the transition relation of a system, while employing \smt-based
  procedures, such as interpolation, to mitigate the limitations of local
  reasoning and allow for better generalization. Unfortunately, these
  mitigations intertwine model checking with heuristics of the underlying
  \smt-solver, negatively affecting stability of model checking.

  In this paper, we propose to tackle the limitations of locality in a
  systematic manner. We introduce explicit \emph{global guidance} into the local
  reasoning performed by \Ict-style algorithms. To this end, we extend the
  \smt-\Ict paradigm with three novel rules, designed to mitigate fundamental
  sources of failure that stem from locality. We instantiate these rules for the
  theory of Linear Integer Arithmetic and implement them on top of \Spacer
  solver in Z3. Our empirical results show that \GSpacer, \Spacer extended with global
  guidance, is significantly more effective than both \Spacer and sole global
  reasoning, and, furthermore, is insensitive to interpolation.

\end{abstract}

 \section{Introduction}
\label{sec:intro}

SMT-based Model Checking algorithms that combine SMT-based search for bounded
counterexamples with interpolation-based search for inductive invariants are
currently the most effective techniques for verification of infinite state systems.
They are widely applicable, including for verification of synchronous systems,
protocols, parameterized systems, and software.

The Achilles heel of these approaches is the mismatch between the \emph{local}
reasoning used to establish absence of bounded counterexamples and a
\emph{global} reason for absence of unbounded counterexamples (i.e., existence
of an inductive invariant). This is particularly apparent in IC3-style
algorithms~\cite{DBLP:conf/vmcai/Bradley11}, such as
\Spacer~\cite{DBLP:conf/cav/KomuravelliGC14}. IC3-style algorithms
establish bounded safety by repeatedly computing predecessors of error (or bad)
states, blocking them by local reasoning about a single step of the transition
relation of the system, and, later, using the resulting \emph{lemmas} to
construct a candidate inductive invariant for the global safety proof. The whole
process is driven by the choice of local lemmas. Good lemmas lead to quick
convergence, bad lemmas make even simple-looking problems difficult to solve.

The effect of local reasoning is somewhat mitigated by the use of interpolation
in lemma construction. In addition to the usual inductive generalization by
dropping literals from a blocked bad state, interpolation is used to further
generalize the blocked state using theory-aware reasoning. For example, when
blocking a bad state $x = 1 \land y = 1$, inductive generalization would infer a
sub-clause of $x \neq 1 \lor y \neq 1$ as a lemma, while interpolation might
infer $x \neq y$ -- a predicate that might be required for the inductive
invariant. \Spacer, that is based on this idea, is extremely effective, as
demonstrated by its performance in recent CHC-COMP competitions~\cite{chc-comp}.
The downside, however, is that the approach leads to a highly unstable procedure
that is extremely sensitive to syntactic changes in the system description,
changes in interpolation algorithms, and any algorithmic changes in the
underlying SMT-solver. 

An alternative approach, often called \emph{invariant inference}, is to focus on
the global safety proof, i.e., an inductive invariant. This has long been
advocated by such approaches as Houdini~\cite{DBLP:conf/fm/FlanaganL01}, and,
more recently, by a variety of machine-learning inspired techniques, e.g.,
FreqHorn~\cite{DBLP:conf/fmcad/FedyukovichKB17},
LinearArbitrary~\cite{DBLP:conf/pldi/ZhuMJ18}, and
ICE-DT~\cite{DBLP:conf/popl/0001NMR16}. The key idea is to iteratively generate
positive (i.e., reachable states) and negative (i.e., states that reach an
error) examples and to compute a candidate invariant that separates these two
sets. The reasoning is more focused towards the invariant, and, the search is
restricted by either predicates, templates, grammars, or some combination.
Invariant inference approaches are particularly good at finding simple inductive
invariants. However, they do not generalize well to a wide variety of problems.
In practice, they are often used to complement other SMT-based techniques.

In this paper, we present a novel approach that extends, what we call,
\emph{local reasoning} of IC3-style algorithms with \emph{global guidance}
inspired by the invariant inference algorithms described above. Our main insight
is that the set of lemmas maintained by IC3-style algorithms hint towards a
potential global proof. However, these hints are lost in existing approaches. We
observe that letting the current set of lemmas, that represent candidate global
invariants, guide local reasoning by introducing new lemmas and states to be
blocked is often sufficient to direct IC3 towards a better global proof.

We present and implement our results in the context of \Spacer\ --- a solver for
Constrained Horn Clauses (CHC) --- implemented in the
Z3 SMT-solver~\cite{DBLP:conf/tacas/MouraB08}. \Spacer is used by multiple
software model checking tools, performed remarkably well in CHC-COMP
competitions~\cite{chc-comp}, and is open-sourced. However, our results are
fundamental and apply to any other IC3-style algorithm. While our implementation
works with arbitrary CHC instances, we simplify the presentation by focusing on
infinite state model checking of transition systems.

\newbox\mrg
\begin{lrbox}{\mrg}
\begin{lstlisting}[language=C++,linewidth=0.4\textwidth]
a, c := 0, 0;
// b, d := a, c; |\label{ln:init-eq}|
b, d := 0, 0; |\label{ln:init-zero}|
while(nd()) 
// inv: a - c = b - d;
{
  if(nd()) { a++; b++; }
  else { c++; d++; }
}
assert(a <=  c => b <= d);
\end{lstlisting}
\end{lrbox}

\newbox \splt
\begin{lrbox}{\splt}
\begin{lstlisting}[language=C++,linewidth=0.4\textwidth, numbers=none]
a, b := 0, 0;
while(nd()) 
// inv: a >= 0 and b >= 0;
{
  a := a + b;
  b++;
}
assert(a >= 0);

|$\phantom{.}$
\end{lstlisting}
\end{lrbox}

\newbox \abs
\begin{lrbox}{\abs}
\begin{lstlisting}[language=C++,linewidth=0.4\textwidth, numbers=none, frame =
none, boxpos = t]
a, b, c := 0, 0, 0;
while(nd()) 
// inv: b = c;
{
  a++; b++; c++;
}
assert(a >= 100 => b = c);


|$\phantom{.}$
\end{lstlisting}
\end{lrbox}

\begin{figure}[t]
  \centering
  \subcaptionbox{myopic generalization\hfill\label{ex:mrg}}{\begin{minipage}{.3\textwidth}\usebox\mrg\end{minipage}}
  \hfill
  \subcaptionbox{excessive \makebox[0pt][l]{generalization}\hfill\label{ex:splt}}{\begin{minipage}{.3\textwidth}\usebox\splt\end{minipage}} 
  \hfill
  \subcaptionbox{stuck in a rut\hfill\label{ex:conjecture}}{\begin{minipage}{.3\textwidth}\usebox\abs\end{minipage}}
  \vspace{-0.1in}
  \caption{Verification tasks to illustrate sources of divergence
    for \Spacer. The call $nd()$ non-deterministically returns a Boolean value.}
  \label{fig:example}
\end{figure}

We illustrate the pitfalls of local reasoning using three examples shown in
\Cref{fig:example}. All three examples are small, simple, and have simple
inductive invariants. All three are challenging for \Spacer. Where these
examples are based on \Spacer-specific design choices, each exhibits a
fundamental deficiency that stems from local reasoning. We believe they can be
adapted for any other IC3-style verification algorithm. The examples assume
basic familiarity with the IC3 paradigm. Readers who are not familiar with it
may find it useful to read the examples after reading \Cref{sec:back}.

\paragraph{Myopic generalization.}
\Spacer diverges on the example in~\Cref{ex:mrg} by iteratively learning
lemmas of the form $( a - c \leq k ) \limp ( b - d \leq k )$ for different
values of $k$, where $a$, $b$, $c$, $d$ are the program variables. These
lemmas establish that there are no counterexamples of longer and longer
lengths. However, the process never converges to the desired lemma $( a - c )
\leq ( b - d )$, which excludes counterexamples of any length.
The lemmas are discovered using interpolation, based on proofs found by the
SMT-solver. A close examination of the corresponding proofs shows that the
relationship between $(a-c)$ and $(b-d)$ does not appear in the proofs, making
it impossible to find the desired lemma by tweaking local interpolation
reasoning. On the other hand, looking at the global proof (i.e., the set of
lemmas discovered to refute a bounded counterexample), it is almost obvious that
$(a-c) \leq (b-d)$ is an interesting generalization to try.
Amusingly, a small, syntactic, but semantic preserving change of
swapping~\cref{ln:init-eq} for \cref{ln:init-zero} in~\Cref{ex:mrg} changes the
SMT-solver proofs, affects local interpolation,  and makes the instance trivial
for \Spacer.

\paragraph{Excessive (predecessor) generalization.}
\Spacer diverges on the example in~\Cref{ex:splt} by computing an infinite
sequence of lemmas of the form $a + k_1 \times b \geq k_2$, where $a$ and $b$
are program variables, and $k_1$ and $k_2$ are integers. The root cause is
excessive generalization in predecessor computation. The $\Bad$ states are $a <
0$, and their predecessors are states such as $(a = 1 \land b = -10)$, $(a = 2
\land b = -10)$, etc., or, more generally, regions $(a + b < 0)$, $(a + 2b <
-1)$, etc. \Spacer always attempts to compute the most general predecessor
states. This is the best local strategy, but blocking these regions by
learning their negation leads to the aforementioned lemmas. According to the
global proof these lemmas do not converge to a linear invariant. An alternative
strategy that under-approximates the problematic regions by (numerically)
simpler regions and, as a result, learns simpler lemmas is desired (and is
effective on this example). For example, region $a + 3b \leq -4$ can be
under-approximated by $a \leq 32 \land b \leq -12$, eventually leading to a
lemma $b \geq 0$, that is a part of the final invariant: $(a \geq 0 \land b \geq
0)$.

\paragraph{Stuck in a rut.}
Finally, \Spacer converges on the example
in~\Cref{ex:conjecture}, but only after unrolling the system for $100$
iterations. During the first $100$ iterations, \Spacer learns that program
states with $(a \geq 100 \land b \neq c)$ are not reachable because $a$ is
bounded by $1$ in the first iteration, by $2$ in the second, and so on. In each
iteration, the global proof is updated by replacing a lemma of the form $a < k$
by lemma of the form $ a < (k+1)$ for different values of $k$. Again, the
strategy is good locally -- total number of lemmas does not grow and the bounded
proof is improved. Yet, globally, it is clear that no progress is made since the
same set of bad states are blocked again and again in slightly different ways.
An alternative strategy is to abstract the literal $a \geq 100$ from the formula
that represents the bad states, and, instead, conjecture that no states in $b
\neq c$ are reachable.

\paragraph{Our approach: global guidance.}
As shown in the examples above, in all the cases that \Spacer diverges, the
missteps are not obvious locally, but are clear when the overall proof is
considered. We propose three new rules, \texttt{Subsume}, \texttt{Concretize},
and, \texttt{Conjecture}, that provide global guidance, by considering existing
lemmas, to mitigate the problems illustrated above. \texttt{Subsume} introduces
a lemma that generalizes existing ones, \texttt{Concretize} under-approximates
partially-blocked predecessors to focus on repeatedly unblocked regions, and
\texttt{Conjecture} over-approximates a predecessor by abstracting away regions
that are repeatedly blocked.
The rules are generic, and apply to arbitrary SMT
theories. Furthermore, we propose an efficient instantiation of the rules for
the theory Linear Integer Arithmetic.

We have implemented the new strategy, called \GSpacer, in \Spacer and compared
it to the original implementation of \Spacer. We show that \GSpacer outperforms
\Spacer in benchmarks from CHC-COMP 2018 and 2019. More significantly, we show
that the performance is independent of interpolation. While \Spacer is highly
dependent on interpolation parameters, and performs poorly when interpolation is
disabled, the results of \GSpacer are virtually unaffected by interpolation. We
also compare \GSpacer to LinearArbitrary~\cite{DBLP:conf/pldi/ZhuMJ18}, a tool
that \emph{infers invariants} using global reasoning.
\GSpacer outperforms LinearArbitrary on the benchmarks
from~\cite{DBLP:conf/pldi/ZhuMJ18}. These results indicate that global guidance
mitigates the shortcomings of local
reasoning.

The rest of the paper is structured as follows. \Cref{sec:back} presents the
necessary background. \Cref{sec:ddalg} introduces our \emph{global guidance} as
a set of abstract inference rules. \Cref{sec:implementation} describes an
instantiation of the rules to Linear Integer Arithmetic (LIA). \Cref{sec:eval}
presents our empirical evaluation. Finally, \Cref{sec:concl} describes related
work and concludes the paper.

 \section{Background}
\label{sec:back}

\paragraph{Logic.} We consider first order logic modulo theories,
and adopt the standard notation and terminology.
A first-order language modulo theory $\cT$ is defined over a signature $\Sig$ that consists of constant,
function and predicate symbols, some of which may be \emph{interpreted} by $\cT$.
As always, \emph{terms} are constant symbols, variables, or function symbols
applied to terms; \emph{atoms} are predicate symbols applied to terms;
\emph{literals} are atoms or their negations; \emph{cubes} are conjunctions of
literals; and \emph{clauses} are disjunctions of literals. Unless otherwise
stated, we only consider \emph{closed} formulas (i.e., formulas without any free
variables). As usual, we use sets of formulas and their conjunctions
interchangeably.

\paragraph{MBP.} Given a set of constants $\Vars$, a formula $\varphi$ and a model $M \models
\varphi$, Model Based Projection (MBP) of $\varphi$ over the constants $\Vars$, denoted
$\mbp(\Vars, \varphi, M)$, computes a model-preserving under-approximation of $\varphi$ projected onto $\Sig \setminus \Vars$.
That is,  $\mbp(\Vars, \varphi, M)$ is a formula over $\Sig \setminus \Vars$ such that
$M \models \mbp(\Vars, \varphi, M)$ and any model $M' \models \mbp(\Vars, \varphi, M)$ can be extended to
a model $M'' \models \varphi$ by providing an interpretation for $\Vars$.
There are polynomial time algorithms for computing MBP in Linear
Arithmetic~\cite{DBLP:conf/cav/KomuravelliGC14,DBLP:conf/lpar/BjornerJ15}. 

\paragraph{Interpolation.} Given an unsatisfiable formula $A \land B$, an
interpolant, denoted $\itp(A,B)$, is a formula $I$ over the shared signature of
$A$ and $B$ such that $A \limp I$ and $I \limp \neg B$.

\paragraph{Safety problem.} A \emph{transition system} is a pair $\langle \Init,
\Tr \rangle$, where $\Init$ is a formula over $\Sig$ and $\Tr$ is a formula over
$\Sig \cup \Sig'$, where $\Sig' = \{s' \mid s \in \Sig\}$.\footnote{In fact, a
  primed copy is introduced in $\Sig'$ only for the uninterpreted symbols in
  $\Sig$. Interpreted symbols remain the same in $\Sig'$.} The states of the system correspond
to structures over $\Sig$, $\Init$ represents the initial states and $\Tr$
represents the transition relation, where $\Sig$ is used to represent the
pre-state of a transition, and $\Sig'$ is used to represent the post-state. For
a formula $\varphi$ over $\Sig$, we denote by $\varphi'$ the formula obtained by
substituting each $s \in \Sig$ by $s' \in \Sig'$.
A \emph{safety problem} is a triple
$\langle \Init, \Tr, \Bad \rangle$, where
$\langle \Init, \Tr \rangle$ is a transition system and $\Bad$ is a
formula over $\Sig$ representing a set of
bad states.

The safety problem $\langle \Init, \Tr, \Bad \rangle$ has a \emph{counterexample
  of length $k$} if the following formula is satisfiable: $ \Init^0 \wedge
\bigwedge_{i=0}^{k-1} \Tr^i \wedge \Bad^k, $ where $\varphi^i$ is defined over
$\Sig^i = \{s^i \mid s \in \Sig\}$ (a copy of the signature used to represent
the state of the system after the execution of $i$ steps) and is obtained from
$\varphi$ by substituting each $s \in \Sig$ by $s^i \in \Sig^i$, and $\Tr^i$ is
obtained from $\Tr$ by substituting $s \in \Sig$ by $s^i \in \Sig^i$ and $s' \in
\Sig'$ by $s^{i+1} \in \Sig^{i+1}$. The transition system is \emph{safe} if the
safety problem has no counterexample, of any length.

\paragraph{Inductive invariants.}
An \emph{inductive invariant} is a formula $\Inv$ over $\Sig$ such that
\begin{inparaenum}[(i)]
\item $\Init \limp \Inv$,
\item $\Inv \wedge \Tr \limp \Inv'$, and
\item $\Inv\limp \neg \Bad$.
\end{inparaenum}
If such an inductive invariant exists, then the transition system is safe.

\paragraph{Spacer.}
The safety problem defined above is an instance of a more general problem,
CHC-SAT, of satisfiability of Constrained Horn Clauses (CHC). \Spacer is a
semi-decision procedure for CHC-SAT. However, to simplify the presentation, we
describe the algorithm only for the particular case of the safety problem. We stress that
\Spacer, as well as the developments of this paper, apply to the more general
setting of CHCs (both linear and non-linear). We assume that the only
uninterpreted symbols in $\Sig$ are constant symbols, which we denote $\Consts$.
Typically, these represent program variables. Without loss of generality, we
assume that $\Bad$ is a cube.

\Cref{alg:spc} presents the key ingredients of \Spacer as a set of guarded
commands (or rules). It maintains the following. Current unrolling depth $N$ at
which a counterexample is searched (there are no counterexamples with depth less
than $N$). A \emph{trace} $\cO = (\cO_0, \cO_1, \ldots)$ of \emph{frames}, such that
each frame $\cO_i$ is a set of \emph{lemmas}, and each lemma $\ell \in \cO_i$ is
a clause.
 A queue of \emph{proof obligations} $Q$, where each proof obligation
(\pob) in $Q$ is a pair $\langle \varphi, i \rangle$ of a cube $\varphi$ and a
level number $i$, $0 \leq i \leq N$. An under-approximation $\cU$ of reachable
states. Intuitively, each frame $\cO_i$ is a candidate inductive invariant s.t.
$\cO_i$ over-approximates states reachable up to $i$ steps from $\Init$.
The latter is ensured since $\cO_0 = \Init$, the trace is monotone, i.e.,
$\cO_{i+1} \subseteq \cO_i$, and each frame is inductive \emph{relative} to its previous one,
i.e., $\cO_i \wedge \Tr \limp \cO_{i+1}'$.
Each \pob $\langle \varphi, i \rangle$ in $Q$ corresponds to a suffix of a potential
counterexample that has to be blocked in $\cO_i$, i.e., has to be proven unreachable in $i$ steps.

The \texttt{Candidate} rule adds an initial \pob $\langle \Bad, N \rangle$ to
the queue. If a \pob $\langle \varphi, i \rangle$ cannot be blocked because
$\varphi$ is reachable from frame $(i-1)$, the \texttt{Predecessor} rule
generates a predecessor $\psi$ of $\varphi$ using MBP and adds $\langle \psi,
i-1 \rangle$ to $Q$. The \texttt{Successor} rule updates the set of reachable
states if the \pob is reachable. If the \pob is blocked, the \texttt{Conflict}
rule strengthens the trace $\cO$ by using interpolation to learn a new lemma
$\ell$ that blocks the \pob, i.e., $\ell$ implies $\neg \varphi$. The
\texttt{Induction} rule strengthens a lemma by inductive generalization and the
\texttt{Propagate} rule pushes a lemma to a higher frame. If the $\Bad$ state
has been blocked at $N$, the \texttt{Unfold} rule increments the depth of
unrolling $N$. In practice, the rules are scheduled to ensure progress towards
finding a counterexample.\DecMargin{0.1in}
\begin{algorithm2e}[t]
  \LinesNotNumbered
  \LinesNumberedHidden
\SetKwIF{Guard}{Guard1}{Guard2}{$\llbracket$}{$\rrbracket$}{}{}{}
  \SetKwComment{Rule}{}{}
  \SetKwFor{select}{forever do}{}{}
  \LinesNotNumbered
  \Fn{\Spacer}{
    \Indm
    \KwIn{$\langle \Init, \Tr, \Bad \rangle$}
    \KwOut{$\langle \safe, \Inv \rangle$ or \unsafe}
    $Q := \emptyset$ \tcp*{\pob queue}
    $N := 0$ \tcp*{maximum safe level}
    $\cO_0 := \Init, \cO_i := \top \textbf{ for all } i > 0$ \tcp*{lemma trace}
    $\cU := \Init$ \tcp*{reachable states}
    \select{}{
      \Rule*[h]{Candidate} \lGuard {$\isSat(\cO_N \land \Bad)$} {$Q := Q \cup
        \langle \Bad, N \rangle$}

      \Rule*[h]{Predecessor} \lGuard {$\langle \varphi, i + 1 \rangle \in Q$, $M \models
        \cO_i \land \Tr \land \varphi'$}
      {$Q := Q \cup \langle \mbp(\Consts', \Tr \land \varphi', M), i \rangle$}

      \Rule*[h]{Successor} \lGuard {$\langle \varphi, i + 1 \rangle \in Q$, $M \models \cF(\cU) \land
        \varphi'$}{$\cU := \cU \lor  \mbp(\Consts, \cF(\cU)
        , M)[\Consts' \mapsto\Consts]$}

      \Rule*[h]{Conflict} \lGuard {$\langle \varphi, i + 1 \rangle \in Q$, $\cF(\cO_i) \limp \neg \varphi'$}
      {$\cO_j := ( \cO_j \land  \itp(\cF(\cO_i), \varphi')[\Consts' \mapsto \Consts] )
        \makebox[0pt][l]{ $\textbf{ for all } j \leq i + 1 $}$}

      \Rule*[h]{Induction} \lGuard{$\ell \in \cO_{i+1}, \ell = (\varphi \lor \psi), \cF(\varphi \land \cO_i) \limp
        \varphi'$}{$\cO_{j} \gets \cO_{j} \land \varphi \textbf{ for all } j
        \leq i + 1$}

      \Rule*[h]{Propagate} \lGuard{$\ell \in \cO_{i}, \cO_i \land \Tr \limp
        \ell'$}{$\cO_{i + 1} \gets (\cO_{i + 1} \land \ell)$}

      \Rule*[h]{Unfold} \lGuard {$\cO_N \limp \neg \Bad$}{$N := N + 1$}

      \Rule*[h]{Safe} \lGuard {$\cO_{i + 1} \limp \cO_{i} {\normalfont \textbf{ for some }} i <
        N$}{\Return $\langle  \safe, \cO_i \rangle$}

      \Rule*[h]{Unsafe} \lGuard {\isSat($\Bad \land \cU$)}{\Return \unsafe}
    }
  }
  \caption{\Spacer algorithm as a set of guarded commands. We use the shorthand
    $\cF(\varphi) = \cU' \lor (\varphi \land \Tr)$. }
  \label{alg:spc}
\end{algorithm2e}
\IncMargin{0.1in}
\DecMargin{0.1in}
\begin{algorithm2e}[t]
  \LinesNotNumbered
\SetKwIF{Guard}{Guard1}{Guard2}{$\llbracket$}{$\rrbracket$}{}{}{}
  \SetKwComment{Rule}{}{}
  \SetKwFor{select}{forever do}{}{}
  \Rule*[h]{Subsume}
  \Guard{$\cL \subseteq \cO_{i}, k\geq i, \cF(\cO_{k}) \limp
    \psi', \forall \ell \in \cL \ldotp \psi \limp \ell$}{

    $\qquad\cO_j \gets (\cO_j \land \psi) \textbf{ for all } j \leq k+1$}

  \vspace{0.07in}
  \Rule*[h]{Concretize}
  \Guard{$\cL \subseteq \cO_i, \langle
    \varphi, j \rangle \in Q, \forall \ell \in \cL \ldotp \isSat(\varphi \land \neg \ell),\isSat(\varphi \land
    \Land \cL), 
    \gamma \limp \varphi, \isSat(\gamma \land \Land \cL)$}
  {$\qquad Q \gets Q \cup \langle \gamma, k + 1 \rangle \textbf{ where } k
    = \max \{ j \mid \cO_j \limp \neg \gamma \} $}

  \vspace{0.07in}
  \Rule*[h]{Conjecture}
  \Guard{$\cL \subseteq \cO_i, \langle \varphi, j \rangle \in Q,
    \varphi \equiv \alpha \land \beta, \forall \ell \in \cL \ldotp \ell \limp
    \neg \beta \land \isSat(\ell \land \alpha), \cU \limp \neg \alpha $}
  {$\qquad Q \gets Q \cup
    \langle \alpha, k + 1 \rangle \textbf{ where } k = \max \{ j \mid \cO_j
    \limp \neg \alpha \}$
      }

  \caption{Global guidance rules for \Spacer.}
  \label{alg:gspacer-rules}
\end{algorithm2e}
\IncMargin{0.1in}

   \section{Global guidance of local proofs}
\label{sec:ddalg}
As illustrated by the examples in \Cref{fig:example}, while \Spacer is generally
effective, its local reasoning is easily confused. The effectiveness is very
dependent on the local computation of predecessors using model-based projection,
and lemmas using interpolation. In this section, we extend \Spacer with three
additional \emph{global} reasoning rules. The rules are inspired by the
deficiencies illustrated by the motivating examples in \Cref{fig:example}. In
this section, we present the rules abstractly, independent of any underlying
theory, focusing on pre- and post-conditions. In \Cref{sec:implementation}, we
specialize the rules for Linear Integer Arithmetic, and show how they are
scheduled with the other rules of \Spacer in an efficient verification
algorithm. The new global rules are summarized in \Cref{alg:gspacer-rules}. We
use the same guarded command notation as in description of \Spacer
in~\Cref{alg:spc}. Note that the rules supplement, and not replace, the ones
in~\Cref{alg:spc}.

\paragraph{Subsume} is the most natural rule to explain. It says that if there
is a set of lemmas $\cL$ at level $i$, and there exists a formula $\psi$ such
that (a) $\psi$ is stronger than every lemma in $\cL$, and (b) $\psi$
over-approximates states reachable in at most $k$ steps, where $k \geq i$, then
$\psi$ can be added to the trace to subsume $\cL$. This rule reduces the size of
the global proof -- that is, the number of total not-subsumed lemmas. Note that
the rule allows $\psi$ to be at a level $k$ that is higher than $i$. The choice
of $\psi$ is left open. The details are likely to be specific to the theory
involved. For example, when instantiated for \lia,
\texttt{Subsume} is sufficient to solve example in~\Cref{ex:mrg}.
Interestingly, \texttt{Subsume} is not likely to be
effective for propositional IC3. In that case, $\psi$ is a clause and the only
way for it to be stronger than $\cL$ is for $\psi$ to be a syntactic
sub-sequence of every lemma in $\cL$, but such $\psi$ is already explored by
local inductive generalization (rule \texttt{Induction} in \Cref{alg:spc}).

\paragraph{Concretize} applies to a \pob, unlike \texttt{Subsume}. It is
motivated by example in \Cref{ex:splt} that highlights the problem of excessive
local generalization. \Spacer always computes as general predecessors as possible. This is necessary
for refutational completeness since in an infinite state system there are
infinitely many potential predecessors. Computing the most general predecessor
ensures that \Spacer finds a counterexample, if it exists. However, this also
forces \Spacer to discover more general, and sometimes more complex, lemmas than
might be necessary for an inductive invariant. Without a global view of the
overall proof, it is hard to determine when the algorithm generalizes too much.
The intuition for \texttt{Concretize} is that generalization is excessive when
there is a single \pob $\langle \varphi, j \rangle$ that is not blocked, yet,
there is a set of lemmas $\cL$ such that every lemma $\ell \in \cL$ partially
blocks $\varphi$. That is, for any $\ell \in \cL$, there is a sub-region
$\varphi_{\ell}$ of \pob $\varphi$ that is blocked by $\ell$ (i.e., $\ell \limp
\neg \varphi_{\ell} $), and there is at least one state $s \in \varphi$ that is
not blocked by any existing lemma in $\cL$ (i.e., $s \models \varphi \land
\bigwedge \cL$). In this case, \texttt{Concretize} computes an
under-approximation $\gamma$ of $\varphi$ that includes some not-yet-blocked
state $s$. The new \pob is added to the lowest level at which $\gamma$ is not
yet blocked. \texttt{Concretize} is useful to solve the example
in~\Cref{ex:splt}.

\paragraph{Conjecture} guides the algorithm away from being stuck in the same
part of the search space. A single \pob $\varphi$ might be blocked by a
different lemma at each level that $\varphi$ appears in. This indicates that the
lemmas are too strong, and cannot be propagated successfully to a higher level.
The goal of the \texttt{Conjecture} rule is to identify such a case to guide the
algorithm to explore alternative proofs with a better potential for
generalization. This is done by abstracting away the part of the \pob that has
been blocked in the past. The pre-condition for \texttt{Conjecture} is the
existence of a \pob $\langle \varphi, j \rangle$ such that $\varphi$ is split
into two (not necessarily disjoint) sets of literals, $\alpha$ and $\beta$.
Second, there must be a set of lemmas $\cL$, at a (typically much lower) level
$i < j$ such that every lemma $\ell \in \cL$ blocks $\varphi$, and, moreover,
blocks $\varphi$ by blocking $\beta$. Intuitively, this implies that while there
are many different lemmas (i.e., all lemmas in $\cL$) that block $\varphi$ at
different levels, all of them correspond to a \emph{local} generalization of
$\neg \beta$ that could not be propagated to block $\varphi$ at higher levels. In this case, \texttt{Conjecture} abstracts
the \pob $\varphi$ into $\alpha$, hoping to generate an alternative way to block
$\varphi$. Of course, $\alpha$ is conjectured only if it is not already blocked
and does not contain any known reachable states.
\texttt{Conjecture} is
necessary for a quick convergence on the example in~\Cref{ex:conjecture}.
In some respect,
\texttt{Conjecture} is akin to widening in Abstract
Interpretation~\cite{DBLP:conf/popl/CousotC77} -- it abstracts a set of states
by dropping constraints that appear to prevent further exploration. Of course,
it is also quite different since it does not guarantee termination. While
\texttt{Conjecture} is applicable to propositional IC3 as well, it is much more
significant in SMT-based setting since in many FOL theories a single literal in
a \pob might result in infinitely many distinct lemmas.

Each of the rules can be applied by itself, but they are most effective in
combination. For example, \texttt{Concretize} creates less general predecessors,
that, in the worst case, lead to many simple lemmas. At the same time,
\texttt{Subsume} combines lemmas together into more complex ones. The
interaction of the two produces lemmas that neither one can produce in
isolation. At the same time, \texttt{Conjecture} helps unstuck the algorithm
from a single unproductive \pob, allowing the other rules to take effect.

 \section{Global guidance for Linear Integer Arithmetic}
\label{sec:implementation}

In this section, we present a specialization of our general rules, shown in
\Cref{alg:gspacer-rules}, to the theory of Linear Integer Arithmetic (\lia).
This requires solving two problems: identifying subsets of lemmas for pre-conditions of the rules (clearly
using all possible subsets is too expensive), and applying the rule once its
pre-condition is met. For lemma selection, we introduce a notion of syntactic
clustering based on anti-unification. For rule application, we exploit basic
properties of \lia for an effective algorithm. Our presentation is focused on \lia
exclusively. However, the rules extend to combinations of \lia with
other theories, such as the combined theory of \lia and
Arrays.

The rest of this section is structured as follows. We begin with a brief
background on \lia in \Cref{sec:liadiv}. We then present our lemma selection
scheme, which is common to all the rules, in \Cref{sec:clstr}, followed by
a description of how the rules \texttt{Subsume} (in \Cref{sec:mrg}),
\texttt{Concretize} (in \Cref{sec:concr}), and \texttt{Conjecture} (in
\Cref{sec:abs}) are instantiated for \lia. We conclude in
\Cref{sec:algo} with an algorithm that integrates all the rules together.

\subsection{Linear Integer Arithmetic: Background}
\label{sec:liadiv}

In the theory of Linear Integer Arithmetic (\lia), formulas are defined over a
signature that includes interpreted function symbols $+$, $-$, $\times$,
interpreted predicate symbols $<$, $\leq$, $\mid$, interpreted constant symbols
$0,1,2,\ldots$, and uninterpreted constant symbols $a, b,\ldots, x,y,\ldots$. We
write $\ints$ for the set interpreted constant symbols, and call them
\emph{integers}. We use \emph{constants} to refer exclusively to the
uninterpreted constants (these are often called \emph{variables} in \lia
literature). Terms (and accordingly formulas) in \lia are restricted to be
\emph{linear}, that is, multiplication is never applied to two
constants.

We write \liamod for the fragment of \lia that excludes divisiblity ($d \mid h$)
predicates. A literal in \liamod is a linear inequality; a cube is a conjunction
of such inequalities, that is, a polytope. We find it convenient to use
matrix-based notation for representing cubes in \liamod. A ground cube $c \in
\liamod$ with $p$ inequalities (literals) over $k$ (uninterpreted) constants is
written as $A \cdot \Consts \leq \Nums$, where $A$ is a $p \times k$ matrix of
coefficients in $\ints^{p \times k}$, $\Consts = (x_1 \cdots  x_k)^T$ is a column vector that consists of the
(uninterpreted) constants, and $\Nums = ( n_1 \cdots n_p)^T$ is a column vector in $\ints^p$. For example, the
cube $x \geq 2 \land 2x + y \leq 3$ is written as $
\begin{bsmallmatrix} -1 & 0 \\ 2 & 1 \end{bsmallmatrix} \cdot \begin{bsmallmatrix} x \\
y \end{bsmallmatrix} \leq \begin{bsmallmatrix} -\;2 \\ \phantom{-}3\end{bsmallmatrix}.
$ In the sequel, all vectors are column vectors, super-script $T$ denotes
transpose, dot is used for a dot product and $[\Nums_1 ; \Nums_2]$ stands for a
matrix of column vectors $\Nums_1$ and $\Nums_2$.

 \subsection{Lemma selection}
\label{sec:clstr}

A common pre-condition for all of our global rules in Alg.~\ref{alg:gspacer-rules} is the
existence of a subset of lemmas $\cL$ of some frame $\cO_i$. Attempting to apply
the rules for every subset of $\cO_i$ is infeasible. In practice, we use
syntactic similarity between lemmas as a predictor that one of the global rules
is applicable, and restrict $\cL$ to subsets of syntactically similar lemmas. In
the rest of this section, we formally define what we mean by \emph{syntactic
  similarity}, and how syntactically similar subsets of lemmas, called
\emph{clusters}, are maintained efficiently throughout the algorithm.

\paragraph{Syntactic similarity.}
A formula $\pi$ with free variables is called a \emph{pattern}. Note that we do
not require $\pi$ to be in \lia.
Let $\sigma$ be
a substitution, i.e., a mapping from variables to terms. We write $\pi\sigma$
for the result of replacing all occurrences of free variables in $\pi$ with
their mapping under $\sigma$. A substitution $\sigma$ is called \emph{numeric}
if it maps every variable to an integer, i.e., the range of $\sigma$ is
$\ints$. We say that a formula $\varphi$ \emph{numerically matches} a pattern
$\pi$ iff there exists a numeric substitution $\sigma$ such that $\varphi =
\pi\sigma$. Note that, as usual, the equality is syntactic. For example,
consider the pattern $\pi = v_0a + v_1b \leq 0$ with free variables $v_0$ and
$v_1$ and uninterpreted constants $a$ and $b$. The formula $\varphi_1 = 3a + 4b
\leq 0$ matches $\pi$ via a numeric substitution $\sigma_1 = \{v_0 \mapsto 3,
v_1 \mapsto 4\}$. However, $\varphi_2 = 4b + 3a \leq 0$, while semantically
equivalent to $\varphi_1$, does not match $\pi$. Similarly $\varphi_3 = a + b
\leq 0$ does not match $\pi$ as well.

Matching is extended to patterns in the usual way by
allowing a substitution $\sigma$ to map variables to variables. We say that a
pattern $\pi_1$ is more general than a pattern $\pi_2$ if $\pi_2$ matches
$\pi_1$. A pattern $\pi$ is a \emph{numeric anti-unifier} for a pair of formulas
$\varphi_1$ and $\varphi_2$ if both $\varphi_1$ and $\varphi_2$ match $\pi$
numerically. We write $\nau{\varphi_1}{\varphi_2}$ for a most general numeric
anti-unifier of $\varphi_1$ and $\varphi_2$. We say that two formulas
$\varphi_1$ and $\varphi_2$ are \emph{syntactically similar} if there exists a
numeric anti-unifier between them (i.e., $\nau{\varphi_1}{\varphi_2}$ is
defined). Anti-unification is extended to sets of formulas in the usual way.

\paragraph{Clusters.}
We use anti-unification to define \emph{clusters} of syntactically similar
formulas. Let $\Phi$ be a fixed set of formulas, and $\pi$ a pattern. A
\emph{cluster}, $\cC_{\Phi}(\pi)$, is a subset of $\Phi$ such that every formula
$\varphi \in \cC_{\Phi}(\pi)$ numerically matches $\pi$. That is,
$\pi$ is a numeric anti-unifier for $\cC_{\Phi}(\pi)$. In the implementation, we restrict the pre-conditions of the global rules so
that a subset of lemmas $\cL \subseteq \cO_i$ is a cluster for some pattern
$\pi$, i.e., $\cL = \cC_{\cO_i}(\pi)$.

\newcommand{\ellnew}{\ell_{\text{new}}}
\paragraph{Clustering lemmas.}
We use the following strategy to efficiently keep track of available clusters.
Let $\ellnew$ be a new lemma to be added to $\cO_i$. Assume there is at least
one lemma $\ell \in \cO_i$ that numerically anti-unifies with $\ellnew$ via some
pattern $\pi$. If such an $\ell$ does not belong to any cluster, a new cluster
$\cC_{\cO_i}(\pi) = \{\ellnew, \ell\}$ is formed, where
$\pi = \nau{\ellnew}{\ell}$. Otherwise, for every lemma
$\ell \in \cO_{i}$ that numerically matches $\ellnew$ and every cluster
$\cC_{\cO_i}(\hat{\pi})$ containing $\ell$, $\ellnew$ is added to
$\cC_{\cO_i}(\hat{\pi})$ if $\ellnew$ matches $\hat{\pi}$, or a new cluster is
formed using $\ell$, $\ellnew$, and any other lemmas in $\cC_{\cO_i}(\hat{\pi})$
that anti-unify with them. Note that a new lemma $\ellnew$ might belong to
multiple clusters.

For example, suppose $\ellnew= (a \leq 6 \lor b \leq 6)$, and there is already a cluster
  $\cC_{\cO_i}(a \leq v_0 \lor b \leq 5) = \{ (a \leq 5 \lor b \leq 5), (a \leq
  8 \lor b \leq 5)\}$. Since $\ellnew$ anti-unifies with each of the lemmas in
  the cluster, but does not match the pattern $a \leq v_0 \lor b \leq 5$, a new
  cluster that includes all of them is formed w.r.t. a more general pattern:
  $\cC_{\cO_i}(a \leq v_0 \lor b \leq v_1) = \{ (a \leq 6 \lor b \leq 6), (a
  \leq 5 \lor b \leq 5), (a \leq 8 \lor b \leq 5)\}$.

In the presentation above, we assumed that anti-unification is completely
syntactic. This is problematic in practice since it significantly limits the
applicability of the global rules. Recall, for example, that $a + b \leq 0$ and
$2a + 2b \leq 0$ do not anti-unify numerically according
to our definitions, and, therefore, do not cluster together. In practice, we
augment syntactic anti-unification with simple rewrite rules that are applied
greedily. For example, we normalize all $\lia$ terms, take care of implicit
multiplication by $1$, and of associativity and commutativity of addition. In
the future, it is interesting to explore how advanced anti-unification
algorithms, such
as~\cite{DBLP:conf/ershov/BulychevKZ09,DBLP:journals/tplp/YernauxV19}, can be
adapted for our purpose.

 \subsection{Subsume rule for LIA}
\label{sec:mrg}
Recall that the \texttt{Subsume} rule~(\Cref{alg:gspacer-rules}) takes a cluster
of lemmas $\cL = \cC_{\cO_i}(\pi)$ and computes a new lemma $\psi$ that subsumes
all the lemmas in $\cL$, that is $\psi \limp \Land \cL$. We find it convenient
to dualize the problem. Let $\cS = \{ \neg \ell \mid \ell \in \cL\}$ be the dual
of $\cL$, clearly $\psi \limp \Land \cL$ iff $(\Lor \cS) \limp \neg \psi$. Note
that $\cL$ is a set of clauses, $\cS$ is a set of cubes, $\psi$ is a clause, and
$\neg \psi$ is a cube. In the case of \liamod, this means that $\Lor \cS$
represents a union of convex sets, and $\neg \psi$ represents a convex set that
the \texttt{Subsume} rule must find. The strongest such $\neg \psi$ in \liamod
exists, and is the convex closure of $\cS$. Thus, applying \texttt{Subsume} in
the context of \liamod is reduced to computing a convex closure of a set of
(negated) lemmas in a cluster.
Full \lia extends \liamod with divisibility
    constraints. Therefore, \texttt{Subsume} obtains a stronger $\neg \psi$ by adding such
    constraints.

    \begin{example}
      For example, consider the following cluster:
      \label{ex:subs}
    \begin{align*}
\cL &= \{ (x > 2 \lor x < 2 \lor y > 3), (x > 4 \lor x < 4 \lor y > 5), (x > 8 \lor x < 8 \lor  y > 9) \} \\
\cS &= \{(x \leq 2 \land x \geq 2 \land y \leq 3), (x \geq 4 \land x \leq 4 \land y \leq 5), (x \geq 8 \land x \leq 8 \land y
      \leq 9) \}
    \end{align*}
The convex closure of $\cS$ in \liamod is $2 \leq x \leq 8 \land y \leq x + 1$.
    However, a stronger over-approximation exists in \lia: $2 \leq x \leq 8 \land y \leq
    x + 1 \land (2 \mid x)$.
    \qed
  \end{example}

In the sequel, we describe \textsc{subsumeCube}~(\Cref{alg:mrg}) which computes a cube $\varphi$ that over-approximates
$(\Lor \cS)$. \texttt{Subsume} is then implemented by removing from $\cL$ lemmas that
are already subsumed by existing lemmas in $\cL$, dualizing the result into $\cS$, invoking \textsc{subsumeCube}
on $\cS$ and returning $\neg \varphi$ as a lemma that subsumes $\cL$.

    Recall that \texttt{Subsume} is tried only in the case $\cL =
    \cC_{\cO_i}(\pi)$. We further require that the negated pattern, $\neg \pi$, is
of the form $A \cdot \Consts \leq \Vars$, where $A$ is a coefficients matrix, $\Consts$ is a vector of constants
    and $\Vars = (v_1 \cdots v_p)^T$ is a vector of $p$ free variables.
    Under this assumption, $\cS$ (the dual of
    $\cL$) is of the form $\{ (A \cdot\Consts \leq \Nums_i) \mid 1 \leq i \leq q\}$,
    where $q = |\cS|$, and for each $1 \leq i \leq q$, $\Nums_i$ is a numeric
    substitution to $\Vars$ from which one of the negated lemmas in $\cS$ is obtained.
    That is, $|\Nums_i| = |\Vars|$.
In \Cref{ex:subs}, $\neg \pi = x \leq  v_1 \wedge -x \leq  v_2 \wedge y \leq v_3$ and
    \begin{align*}
  A &= \begin{bmatrix} \phantom{-}1 & \phantom{-}0 \\ -1 & \phantom{-}0 \\ \phantom{-}0 & \phantom{-}1 \end{bmatrix} &
   \Consts = \begin{bmatrix} x \\ y \end{bmatrix} \quad
   \Vars = \begin{bmatrix} v_1 \\ v_2  \\ v_3 \end{bmatrix} \quad
   \Nums_1 = \begin{bmatrix} \phantom{-}2 \\ -2 \\ \phantom{-}3 \end{bmatrix}  \quad
  \Nums_2 = \begin{bmatrix}\phantom{-} 4 \\ -4 \\ \phantom{-}5 \end{bmatrix} \quad
  \Nums_3 = \begin{bmatrix} \phantom{-}8 \\ -8 \\ \phantom{-}9 \end{bmatrix}
     \end{align*}

    Each cube $(A \cdot \Consts \leq \Nums_i )\in \cS$ is
    equivalent to $\exists \Vars \ldotp A \cdot \Consts \leq \Vars \land (\Vars
    = \Nums_i)$.
Finally, $( \Lor \cS ) \equiv \exists \Vars \ldotp (A \cdot \Consts
    \leq \Vars) \land (\Lor (\Vars = \Nums_i))$. Thus, computing the
over-approximation of $\cS$ is reduced to (a) computing the convex hull $H$ of a set of
    points $\{\Nums_i \mid 1 \leq i \leq q\}$, (b) computing divisibility constraints
    $D$  that are satisfied by all the points, (c) substituting $H \wedge D$ for the
    disjunction in the equation above, and (c) eliminating variables $\Vars$.
Both the computation of $H \land D$ and the elimination of $\Vars$ may be
    prohibitively expensive. We, therefore, over-approximate them.
Our approach for doing so
    is presented in \Cref{alg:mrg}, and explained in detail below.

\paragraph{Computing the convex hull of $\{\Nums_i \mid 1 \leq i \leq q\}$.}
\crefrange{ln:kr}{ln:scc} compute the convex hull of $\{\Nums_i \mid 1 \leq i \leq q\}$
as a formula over $\Vars$, where variable $v_j$, for $1\leq j \leq p$, represents the $\coord{j}$
coordinates in the vectors (points) $\Nums_i$. Some of the coordinates, $v_j$, in these vectors may
be linearly dependent upon others. To simplify the problem, we first identify
such dependencies and compute a set of linear equalities that expresses them
($L$ in line~\ref{ln:eq}). To do so, we consider a matrix $N_{q \times p}$, where the
$\coord{i}$ row consists of $\Nums_i^T$. The $\coord{j}$ column in $N$, denoted $N_{*j}$,
corresponds to the $\coord{j}$ coordinate, $v_j$. The rank of $N$ is the number of
linearly independent columns (and rows). The other columns (coordinates) can be
expressed by linear combinations of the linearly independent ones. To compute
these linear combinations we use the kernel of $[N ; \vec{1}]$ ($N$ appended
with a column vector of $1$'s), which is the set of all vectors $\vec{y}$ such
that $[N ; \vec{1}]\cdot\vec{y} = \vec{0}$, where $\vec{0}$ is the zero vector.
Let $B = \kernel([N ; \vec{1}])$ be a basis for the kernel of $[N ; \vec{1}]$.
Then $|B| = p - \rank(N)$, and
for each vector $\vec{y} \in B$, the linear equality
$[ v_1  \cdots  v_p \; 1] \cdot \vec{y} =
0$ holds in all the rows of $N$ (i.e., all the given vectors satisfy it).
We accumulate these equalities, which capture
the linear dependencies between the coordinates, in $L$. Further, the equalities are used to compute
$\rank(N)$
coordinates (columns in $N$) that are linearly independent
and, modulo $L$, uniquely determine the remaining coordinates.
We denote by $\lind{\Vars}$ the subset of $\Vars$
that consists of the linearly independent coordinates.
We further denote by
$\lind{\Nums_i}$ the projection of $\Nums_i$ to these coordinates and
by $\lind{N}$ the projection of $N$ to the corresponding columns.
We have that $(\Lor (\Vars = \Nums_i)) \equiv L \wedge (\Lor (\lind{\Vars} = \lind{\Nums_i})$.

In \Cref{ex:subs}, the numeral matrix is $N = \begin{bsmallmatrix} 2 & \;-2 & \phantom{-}3 \\ 4 & \;-4 & \phantom{-}5 \\ 8 & \;-8 & \phantom{-}9 \end{bsmallmatrix}$,
for which $\kernel([N ; \vec{1}]) = \{ \begin{psmallmatrix} 1 & 1 &
  0 & 0 \end{psmallmatrix}^T,  \begin{psmallmatrix} \;1 & 0 & \;-1 & 1 \end{psmallmatrix}^T \}$.
Therefore, $L$ is the conjunction of equalities $v_1 + v_2 = 0 \land v_1 - v_3
+ 1 = 0 $, or, equivalently $v_3 =  v_1 + 1 \land v_2 = -v_1$,
$\lind{\Vars} = \begin{pmatrix} v_1 \end{pmatrix}^T$, and
\begin{align*}
\lind{\Nums_1} = \begin{bmatrix} 2 \end{bmatrix} \qquad
\lind{\Nums_2} = \begin{bmatrix} 4 \end{bmatrix} \qquad
\lind{\Nums_3} = \begin{bmatrix} 8 \end{bmatrix} \qquad
\lind{N} = \begin{bmatrix} 2 \\ 4 \\ 8 \end{bmatrix}
\end{align*}

Next, we compute the convex closure of $\Lor (\lind{\Vars} = \lind{\Nums_i})$,
and conjoin it with $L$ to obtain $H$, the convex closure of $(\Lor (\Vars = \Nums_i))$.

If the dimension of $\lind{\Vars}$ is one, as is the case in the example above,
convex closure, $C$, of $\Lor (\lind{\Vars} = \lind{\Nums_i})$ is obtained by bounding
the sole element of $\lind{\Vars}$ based on its values in $\lind{N}$~(line~\ref{ln:bnd}).
In \Cref{ex:subs}, we obtain $C = 2 \leq v_1 \leq 8$.

If the dimension of $\lind{\Vars}$ is greater than one, just computing the
bounds of one of the constants is not sufficient. Instead, we use the concept of
syntactic convex closure from~\cite{DBLP:journals/tplp/BenoyKM05} to compute
the convex closure of $\Lor (\lind{\Vars} = \lind{\Nums_i})$ as $\exists \Alphas
\ldotp C$ where $\Alphas$
is a vector that consists of $q$ fresh \emph{rational} variables and $C$ is
defined as follows (line~\ref{ln:scc}): $C = \Alphas \geq 0 \land \Sigma \Alphas = 1 \land
               \Alphas^T \cdot \lind{N} = (\lind{\Vars})^T$.
$C$ states that $(\lind{\Vars})^T$ is a convex combination of the rows of
$\lind{N}$, or, in other words, $\lind{\Vars}$ is a convex combination of $\{ \lind{\Nums_i}
\mid 1 \leq i \leq q\}$.

To illustrate the syntactic convex closure, consider a second example
with a set of cubes:
$ \cS = \{(x \leq 0 \land y \leq 6),
  (x \leq 6 \land y \leq 0),  (x \leq 5 \land y \leq 5)\}$.
The coefficient matrix $A$, and the numeral matrix $N$ are then:
$A = \begin{bsmallmatrix} 1 & 0 \\ 0 & 1 \end{bsmallmatrix}$ and
$N = \begin{bsmallmatrix} 0 & 6 \\ 6 & 0 \\ 5 & 5\end{bsmallmatrix}$.
Here, $\kernel([N ; \vec{1}])$ is empty -- all the columns
are linearly independent, hence, $L = \true$ and $\lind{\Vars} = \Vars$. Therefore,
syntactic convex closure is applied to the full matrix $N$, resulting in \begin{multline*}
  C = (\alpha_1 \geq 0) \land (\alpha_2 \geq 0) \land (\alpha_3 \geq 0)  \land
         (\alpha_1 + \alpha_2 + \alpha_3 = 1) \land{} \\
         (6\alpha_2 + 5\alpha_3 = v_1) \land (6\alpha_1 + 5\alpha_3 = v_2)
         \end{multline*}
The convex closure of $\Lor ({\Vars} = {\Nums_i})$ is then $L \wedge \exists \Alphas \ldotp C$, which is $\exists \Alphas \ldotp C$ here. 

\paragraph{Divisibility constraints.}
Inductive invariants for verification problems often require divisibility constraints.
We, therefore, use such constraints, denoted $D$, to obtain a stronger
over-approximation of $\Lor ({\Vars} = {\Nums_i})$ than the convex closure. To
add a divisibility constraint for $v_j \in \lind{\Vars}$,
we consider the column $\lind{N_{*j}}$ that corresponds to $v_j$ in $\lind{N}$.
We find the largest positive integer $d$ such that each integer in
$\lind{N_{*j}}$ leaves the same remainder when divided by $d$; namely, there exists $0 \leq r < d$
such that $n \bmod d = r$ for every $n \in \lind{N_{*j}}$.
This means that $d \mid (v_j - r)$ is satisfied by all the points $\Nums_i$.
Note that such $r$ always exists for $d=1$.
To avoid this trivial case, we add the constraint
$d \mid (v_j - r)$ only if $d \neq 1$~(\cref{ln:mod}). We repeat this process for each $v_j \in
\lind{\Vars}$.

In \Cref{ex:subs}, all the elements in the (only) column of the matrix $\lind{N}$, which
corresponds to $v_1$, are divisible by $2$, and no larger $d$ has a corresponding $r$. Thus, \cref{ln:mod} of
\Cref{alg:mrg} adds the divisibility condition $(2 \mid v_1)$ to $D$.

\paragraph{Eliminating existentially quantified variables using MBP.}
By combining the linear equalities exhibited by $N$, the convex closure of
$\lind{N}$ and the divisibility constraints on $\Vars$, we obtain $\exists
\Alphas \ldotp L \wedge C \wedge D$ as an over-approximation of $\Lor ({\Vars} =
{\Nums_i})$. Accordingly, $\exists \Vars \ldotp \exists \Alphas \ldotp \psi$,
where $\psi = (A \cdot \Consts \leq \Vars) \wedge L \wedge C \wedge D$, is an
over-approximation of $(\Lor \cS) \equiv \exists \Vars \ldotp (A \cdot \Consts
\leq \Vars) \land (\Lor (\Vars = \Nums_i))$ (\cref{ln:psi}).
In order to get a \lia cube that overapproximates $\Lor \cS$,
it remains to eliminate the existential quantifiers.
Since quantifier elimination is expensive, and does not necessarily generate convex formulas (cubes),
we approximate it using MBP. Namely, we obtain a cube $\varphi$
that under-approximates $\exists \Vars \ldotp \exists \Alphas \ldotp \psi$
by applying MBP on $\psi$ and a model $M_0 \models \psi$.
We then use an \smt solver to drop
literals from $\varphi$ until it over-approximates $\exists \Vars \ldotp \exists
\Alphas \ldotp \psi$, and hence also $\Lor \cS$~(lines~\ref{ln:ovr-start} to \ref{ln:ovr-end}).
The result is returned by \texttt{Subsume} as an over-approximation of $\Lor \cS$.

Models $M_0$ that satisfy $\psi$ and do not satisfy any of the cubes in $\cS$
are preferred when computing MBP~(line~\ref{ln:mbpmdl})
as they ensure that the result of MBP is not subsumed by any of the cubes in $\cS$.

Note that the $\Alphas$ are rational variables and $\Vars$ are integer
variables, which means we require MBP to support a mixture of integer and
rational variables. To achieve this, we first relax all constants to be
rationals and apply MBP over LRA to eliminate $\Alphas$. We then adjust the
resulting formula back to integer arithmetic by multiplying each atom by the
least common multiple of the denominators of the coefficients in it. Finally, we
apply MBP over the integers to eliminate $\Vars$.

Considering \Cref{ex:subs} again, we get that
$\psi = (x \leq v_1) \land (-x \leq v_2) \land (y \leq v_3) \land (v_3 = 1 +
v_1) \land (v_2 = -v_1) \land (2 \leq
v_1 \leq 8) \land (2 \mid v_1) $ (the first three conjuncts correspond to $(A
\cdot (x\; y)^T \leq (v_1\; v_2 \; v_3)^T)$). Note that in this case we do not have
rational variables $\Alphas$ since $|\lind{\Vars}| = 1$. Depending
on the model, the result of MBP can be one of
\begin{align*}
  y \leq x + 1 \land 2 \leq x \leq 8 \land (2 \mid y - 1) \land (2 \mid x)&\qquad
  x \geq 2 \land x \leq 2 \land y \leq 3 & \\
  y \leq x + 1 \land 2 \leq x \leq 8 \land (2 \mid x)&\qquad
  x \geq 8 \land x \leq 8 \land y \leq 9 & \\
  y \geq x + 1 \land y \leq x + 1 \land 3 \leq y \leq 9 \land (2 \mid y - 1)&\qquad
\end{align*}
However, we prefer a model that does not satisfy any
cube in $\cS = \{(x \geq 2 \land x \leq 2 \land y \leq 3), (x \leq 4 \land x \geq 4 \land y \leq 5), (x \leq 8 \land x \geq 8 \land y
      \leq 9) \}$, rules off the two possibilities on the right. None of these
cubes cover $\psi$, hence generalization is used.

If the first cube is obtained by MBP, it is generalized into $y \leq x + 1 \land
x \geq 2 \land x \leq 8 \land (2 \mid x)$; the second cube is already an
over-approximation; the third cube is generalized
into $y \leq x + 1 \land y \leq  9$. Indeed, each of these
cubes over-approximates $\Lor \cS$.

\begin{figure}[t]
  \LinesNumbered
  \begin{adjustbox}{scale=0.9}
    \begin{minipage}[t]{7.5cm}
      \vspace{0pt}

\begin{algorithm2e}[H]
  \Fn{\textsc{subsumeCube}} {
    \Indm
    \KwIn{$\cS = \{(A \cdot \Consts \leq \Nums_i) \mid 1 \leq i \leq q\}$,
}
    \KwOut{An over-approximation of $(\Lor \cS)$.}

    \tcc{$\Vars$ are integer variables such that: $(\Lor \cS) \iff \exists \Vars
      \ldotp (A\cdot \Consts \leq \Vars) \land (\Lor \Vars = \Nums_i)$}
    $N \gets [\Nums_1 ; \cdots ; \Nums_q ]^T$\;
\tcc{Compute the set of linear dependencies implied by $N$}
    $B \gets \kernel([N ; \vec{1}])$\;\label{ln:kr}
    $L \gets \bigwedge_{\vec{y} \in B} \begin{psmallmatrix} v_1 & \cdots & v_p &
      1\end{psmallmatrix} \cdot \vec{y} = 0$\;\label{ln:eq}
    \If{$|\lind{\Vars}| = 1$} {
      \tcp{Convex closure over a single constant $v_i \in \lind{\Vars}$ }
      $C \gets \min(N_{*i}) \leq v_i \leq \max(N_{*i})$ \label{ln:bnd}
    }
    \Else{
      \tcp{Syntactic convex closure}

$C \gets (\Alphas^T \cdot \lind{N} = (\lind{\Vars})^T) \land (\Sigma \Alphas = 1) \land (\Alphas \geq
      0)$ \;\label{ln:scc}

    }
    \tcc{Compute divisibility constraints}
    $D \gets \top$\;
    \For{$v_j \in \lind{\Vars}$}{
      \If{$\exists d, r \ldotp d \neq 1 \land (\forall n \in \lind{N_{*j}} \ldotp
        (n \bmod d = r))$}{$D \gets D \land d \mid (v_j - r)$  \label{ln:mod}}
    }
$\psi \gets (A\cdot \Consts \leq \Vars) \land L \land C\land D $ \; \label{ln:psi}

    \tcc{Under-approximate quantifier elimination}

    find $M_0$ s.t. $M_0 \models \psi$ and, if possible, $M_0 \not\models (\Lor S)$\label{ln:mbpmdl}\;

    $\varphi \gets \mbp((\Alphas\;\Vars), \psi, M_0)$ \label{ln:mbpcc}\;

    \tcc{Over-approximate quantifier elimination}
    \While{$\isSat(\neg \varphi \land \psi)$} { \label{ln:ovr-start}
      find $M_1$ s.t. $M_1 \models (\neg \varphi \land \psi)$\;
      $\varphi \gets \bigwedge \{\ell \in \varphi \mid  \neg ( M_1 \models \neg \ell )\}$\;
    }

    \Return $\varphi$\label{ln:ovr-end}
    \BlankLine
  }
  \caption{An implementation of the \texttt{Subsume} rule for the dual of a cluster $\cS = \{A\cdot     \Consts \leq \Nums_i \mid 1 \leq i \leq q$\}.}
  \label{alg:mrg}
\end{algorithm2e}
\end{minipage}
\end{adjustbox}
\hfill
\begin{adjustbox}{scale=0.9}
  \begin{minipage}[t]{6.5cm}
    \vspace{0pt}
\begin{algorithm2e}[H]
  \Fn{\concr} {
    \Indm
    \KwIn{A \pob $\langle \varphi, j \rangle$ in \liamod, a cluster of \liamod lemmas $\cL = \cC_{\cO_i}(\pi)$ s.t.\
    $\pi$ is non-linear, $\isSat(\varphi
      \land \Land \cL)$ }
    \KwOut{A cube $\gamma$ such that $\gamma \limp \varphi$ and $\forall \ell
      \in \cL \ldotp\isSat(\gamma
      \land \ell)$}
    $U \gets \{x \mid \coeff{x}{\pi} \in \fVar{\pi}\}$\label{ln:nlu}\;
    find $M$ s.t.\ $M \models \varphi \land \Land \cL$\label{ln:mdlconcr}\;

    $\gamma \gets \top$\;\label{ln:concrstart}
    \ForEach{$\lit \in \varphi$} {
      \lIf{$\const{\lit} \cap U \neq \emptyset$}{$\gamma \gets \gamma \land
        \concrlit(\lit, M, U)$\label{ln:gual}}
      \lElse{$\gamma \gets \gamma \land \lit$\label{ln:remains}}
    }
    $\gamma \gets \textsc{rm\_subsume}(\gamma)$\; \label{ln:concrend}
    \Return{$\gamma$}
  }
  \BlankLine
  \Fn{\concrlit} {
    \Indm
    \KwIn{A literal $\lit = \Sigma_i n_ix_i \leq b_j$ in \liamod, model $M
      \models \lit$, and a set of constants $U$}
    \KwOut{A cube $\gamma^{\lit}$ that concretizes $\lit$}
    \tcc{Construct a single literal using all the constants in {\rm $\const{\lit}\setminus U$}}
    $\gamma^{\lit} \gets \emptyset$\;\label{ln:conclitstart}
    $s \gets 0$ \;
    \ForEach{$x_i \in \const{\lit} \setminus U$} {
      $s  \gets s + n_i x_i$ \label{ln:summands}}
    $\gamma^{\lit} \gets (s \leq M[s])$ \label{ln:sum}\;
    \tcc{Generate one dimensional literals for each constant in $U$}
    \ForEach{$x_i \in \const{\lit} \cap U$} {
      $\gamma^{\lit} \gets \gamma^{\lit} \land (n_ix_i \leq M[n_ix_i])$\label{ln:oned}\;
    }\label{ln:conclitend}
    \Return{$\gamma^{\lit}$}
  }
  \caption{An implementation of the \texttt{Concretize} rule in \lia.\newline \protect\phantom{$\{\cS\}$}}
  \label{alg:gual}
\end{algorithm2e}
\end{minipage}
\end{adjustbox}
\end{figure}
 \subsection{Concretize rule for LIA}
\label{sec:concr}

The \texttt{Concretize} rule~(\Cref{alg:gspacer-rules}) takes a cluster
of lemmas $\cL = \cC_{\cO_i}(\pi)$ and a \pob $\langle \varphi, j \rangle$ such
that each lemma in $\cL$ partially blocks $\varphi$, and creates a new \pob
$\gamma$ that is still not blocked by $\cL$, but $\gamma$ is more concrete,
i.e., $\gamma \limp \varphi$. In our implementation, this rule is applied when
$\varphi$ is in \liamod.
We further require that the pattern, $\pi$, of $\cL$ is non-linear, i.e., some of the constants
appear in $\pi$ with free variables as their coefficients. We denote these constants by $U$.
An example is the pattern $\pi = v_0 x + v_1 y + z \leq 0$, where $U = \{x,y\}$.
Having such a cluster is an indication that attempting to block $\varphi$ in full
with a single lemma may require to track non-linear correlations between the constants,
which is impossible to do in \lia.
In such cases, we identify the coupling of the constants in $U$ in \pobs (and hence in lemmas)
as the potential source of non-linearity.
Hence, we concretize (strengthen) $\varphi$ into a \pob $\gamma$
where the constants in $U$ are no longer coupled to any other constant.

\paragraph{Coupling.}
Formally, constants $u$ and $v$ are \emph{coupled} in a cube $c$, denoted
$\coup{u}{v}{c}$, if there exists a literal $\lit$ in $c$ such that both $u$ and $v$
appear in $\lit$ (i.e., their coefficients in $\lit$ are non-zero).
For example, $x$ and $y$ are
coupled in $x + y \leq 0 \land z \leq 0$ whereas neither of them are coupled
with $z$. A constant $u$ is said to be \emph{isolated} in a cube $c$, denoted
$\iso{u}{c}$, if it appears in $c$ but it is not coupled with any other constant in $c$.
In the above cube, $z$ is isolated.

\paragraph{Concretization by decoupling.} Given a \pob $\varphi$ (a cube) and  a cluster $\cL$,
\Cref{alg:gual} presents our approach for concretizing $\varphi$ by decoupling the
constants in $U$ --- those that have variables as coefficients in the pattern of $\cL$ (line~\ref{ln:nlu}).
Concretization is guided by a model $M \models \varphi \wedge \bigwedge \cL$, representing
a part of $\varphi$ that is not yet blocked by the lemmas in $\cL$ (line~\ref{ln:mdlconcr}).
Given such $M$, we concretize $\varphi$ into a \emph{model-preserving} under-approximation that
isolates all the constants in $U$ and preserves all other couplings. That is, we
find a cube $\gamma$, such that
\begin{align} \label{eq:conc}
  \gamma &\limp \varphi &
   M &\models \gamma &
  \forall u \in U &\ldotp \iso{u}{\gamma} &
  \forall u, v \not \in U \ldotp ( \coup{u}{v}{\varphi} ) &\limp ( \coup{u}{v}{\gamma} )
\end{align}
Note that $\gamma$ is not blocked by $\cL$ since $M$ satisfies both $\bigwedge
\cL$ and $\gamma$. For example, if $\varphi = (x + y \leq 0) \land (x - y \leq
0) \land (x + z \geq 0)$ and $M = [x = 0, y = 0, z = 1]$, then $\gamma = 0 \leq
y \leq 0 \land x \leq 0 \land x + z \geq 1$ is a model preserving
under-approximation that isolates $U = \{y\}$.

\Cref{alg:gual} computes such a cube $\gamma$
by a point-wise concretization of the literals of $\varphi$
followed by the removal of subsumed literals.
Literals that do not contain constants from $U$ remain unchanged. A literal of the form $\lit = t \leq b$, where $t=\sum_i n_ix_i$ (recall that every literal in \liamod can be normalized to this form),
that includes constants from $U$ is concretized into a \emph{cube} by (1)~isolating each of the summands $n_i x_i$ in $t$ that include $U$ from the rest, and (2)~for each of the resulting sub-expressions creating a literal that uses its value in $M$ as a bound. Formally, $t$ is decomposed to $s + \sum_{x_i \in U} n_i x_i$, where $s = \sum_{x_i \not \in U} n_i x_i$.
The concretization of $\lit$ is the cube $\gamma^{\lit} = s \leq M[s] \wedge \bigwedge_{x_i \in U} n_i x_i \leq M[n_i x_i]$, where $M[t']$ denotes the interpretation of $t'$ in $M$.
Note that $\gamma^{\lit}  \limp \lit$ since the bounds are stronger than the original bound on $t$:
$M[s] + \sum_{x_i \in U} M[n_i x_i] = M[t] \leq b$.
This ensures that $\gamma$, obtained by the conjunction of literal concretizations, implies $\varphi$.
It trivially satisfies the other conditions of \Cref{eq:conc}.

For example, the concretization of the literal $(x + y \leq 0)$
with respect to $U = \{y\}$ and $M = [x = 0, y = 0, z = 1]$ is the cube $x \leq 0 \wedge y \leq 0$.
Applying concretization in a similar manner to all the literals
of the cube $\varphi = (x + y \leq 0) \land (x - y \leq 0) \land (x + z \geq 0)$
from the previous example, we obtain the concretization
$ x \leq 0 \land 0 \leq y \leq 0 \land x + z \geq 0$.
Note that the last literal is not concretized as it does not include $y$.

 \subsection{Conjecture rule for LIA}
\label{sec:abs}

The \texttt{Conjecture} rule~(see \Cref{alg:gspacer-rules}) takes a set of
lemmas $\cL$ and a \pob $\varphi \equiv \alpha \land \beta$ such that all lemmas
in $\cL$ block $\beta$, but none of them blocks $\alpha$, where $\alpha$ does
not include any known reachable states. It returns $\alpha$ as a new \pob.

For LIA, \texttt{Conjecture} is applied 
when the following conditions are met: (1)~the \pob $\varphi$ is of the form $\varphi_1 \land \varphi_2
\land \varphi_3$, where $\varphi_3 = (\Nums^T \cdot \Consts \leq b)$, and $\varphi_1$ and
$\varphi_2$ are any cubes. The sub-cube $\varphi_1 \land \varphi_2$ acts as $\alpha$,
while the sub-cube $\varphi_2 \land \varphi_3$ acts as $\beta$. (2)~The cluster $\cL$ consists of $\{bg \lor (\Nums^T
\cdot \Consts \geq b_i) \mid 1 \leq i \leq q\}$, where $b_i > b$ and $bg \limp
\neg \varphi_2$. This means that each of the lemmas in $\cL$ blocks $\beta = \varphi_2 \land \varphi_3$,
and they may be ordered as a sequence of increasingly stronger lemmas,
indicating that they were created by trying to block the \pob at different levels, leading to too strong lemmas that
failed to propagate to higher levels.
(3)~The formula $(bg \lor (\Nums^T \cdot \Consts \geq b_i)) \land \varphi_1 \land
\varphi_2$ is satisfiable, that is, none of the lemmas in $\cL$ block $\alpha =\varphi_1 \land
\varphi_2$, and (4)~$\cU \limp \neg (\varphi_1 \land \varphi_2)$, that is, no state in $\varphi_1 \land \varphi_2$
is known to be reachable. If all four conditions
are met, we conjecture $\alpha = \varphi_1 \land \varphi_2$.
This is implemented by \textsc{conjecture}, that returns $\alpha$ (or $\bot$
when the pre-conditions are not met).

For example, consider the \pob $\varphi = x \geq 10 \land (x + y \geq 10) \land y
\leq 10$ and a cluster of lemmas $\cL = \{ (x + y \leq 0 \lor y \geq 101), (x +
y \leq 0 \lor y \geq 102)\}$. In this case, $\varphi_1 = x \geq 10$, $\varphi_2 = (x + y
\geq 10)$,  $\varphi_3 = y \leq 10$, and $bg =x + y \leq 0$. Each of the lemmas in $\cL$ block $\varphi_2
\land \varphi_3$ but none of them block $\varphi_1 \land \varphi_2$. Therefore, we
conjecture $\varphi_1 \land \varphi_2$: $x \geq 10 \land (x + y \geq
10)$.

 \subsection{Putting it all together}
\label{sec:algo}

\setlength{\textfloatsep}{0pt}
\LinesNumbered
\begin{algorithm2e}[t]
  \caption{\GSpacer for LIA.}
  \label{alg:impl}
  \setlength{\columnsep}{15pt}
  \begin{multicols}{2}

    \AlgoDontDisplayBlockMarkers\SetAlgoNoLine
    \SetKw{Continue}{continue}
    \Fn{\csm} {
      \Indm
      \KwIn{$\langle {\Init, \Tr, \Bad} \rangle$}
      \KwOut{An Inductive invariant or \unsafe}\SetKw{Continue}{continue}
      \Indp
      \tcc{Initialize state of the solver}
      $Q \gets \emptyset; N \gets 0; \cU \gets \Init;$\;
      $ \cO_0 \gets \Init; \cO_i \gets \top, \forall i > 0$\;
      $\textsc{Enqueue}(Q, \langle \Bad, 0 \rangle)$\;
      \While{$\top$}{
        \label{ln:recblk}
        $\langle \varphi, i \rangle \gets \textsc{Pop}(Q)$\;
\If{$\textsc{ConcretizePOB}(\langle \varphi, i \rangle) = \top$} {\label{ln:concrposs}
          \Continue
      }
\If{$\isSat(\cF(\cO_{i-1}) \land \varphi')$}{
        \tcp{The \pob $\varphi$ cannot be blocked at $i$}
        \textsc{AddPredecessor}($\langle {\varphi, i} \rangle$) \label{ln:pred}\;
        \If{$\isSat(\cU \land \Bad)$} {\Return \unsafe \label{ln:unsafe}
          \tcp*[f]{Unsafe}}
      }
      \Else {
        \tcp{The \pob $\varphi$ can be blocked at $i$}
        \textsc{Block}($\langle \varphi, i \rangle$) \label{ln:blk}\;
        \For{$0 \leq j \leq N$}{\For{$\ell \in \cO_j\setminus
            \cO_{j+1}$}{\If{$\cO_j \land \Tr \limp \ell'$}{$\cO_{j + 1} \gets
              \cO_{j + 1} \land \ell$\tcp*[f]{Propagate}}}}

        \If{$\exists 0 \leq j < N \cdot \cO_j \limp \cO_{j - 1}$} {\Return{
            $\langle \safe, \cO_j\rangle$} \label{ln:safe}\tcp*[f]{Safe}}
        \If{$\cO_N \limp \neg \Bad$}{
          $N \gets N + 1$\tcp*{Unfold}
          $\textsc{push}(Q, \langle \Bad, N \rangle)$
        }
        \BlankLine
      }
    }
  }
  \Fn{\textsc{ConcretizePOB}} {
    $\langle \pi_1, \cL_1 \rangle \gets \pobclstr{\varphi}{i}$\label{ln:concrclstr}\;
    $\cL_2 \gets \{\ell \mid \ell \in \cL_1 \land \isSat(\ell \land \varphi)
    \land \isSat(\neg \ell \land \varphi)\}$\label{ln:concrfilter}\;
    \If{$(\cL_2 \neq \emptyset \land \nonLin(\pi_1) \land \isSat(\Land \cL_2 \land \varphi))$} { \label{ln:concrpre}
$\gamma \gets \concr(\varphi, \langle \pi_1, \cL_2\rangle)$\;
      $k \gets \max \{ j \mid \cO_j \limp \neg \gamma \}$\;
      $\textsc{push}(Q, \langle \gamma, k
      \rangle)$\label{ln:concr}\tcp*{Concretize}
      $\textsc{push}(Q, \langle \varphi, i\rangle)$\;
      \Return $\top$
    }
    \lElse {\Return $\bot$}
  }
  \BlankLine
  \Fn{\textsc{AddPredecessor}} {
    \If{$\isSat(\cF(\cU) \land \varphi')$} {
      find $M_1 $ s.t $M_1 \models \cF(\cU) \land \varphi'$\;
      $s \gets (\mbp(\Consts, \cF(\cU), M_1)[\Consts' \mapsto \Consts])$\;
      $\cU \gets \cU \lor s $\label{ln:succ}\tcp*{Successor}
      \Return
    }
    find $M_2 $ s.t $M_2 \models \varTheta$\;
    $p \gets \mbp(\Consts', \Tr \land \varphi', M_2)$\;
    $\textsc{push}(Q, \langle p, i -
    1\rangle)$ \tcp*{Predecessor}
    $\textsc{push}(Q, \langle \varphi, i\rangle)$\;
  }
  \BlankLine
  \Fn{\textsc{Block}}{
       $\ell \gets \gen(\cF(\cO_{i - 1}),
       \varphi')$ \label{ln:indgen} \tcp*{Conflict} \lFor{$0 \leq j \leq i$}{$\cO_j \gets \cO_j \land \ell$}
        $\langle \pi_3, \cL_3 \rangle = \lclstr{\ell}$\;
$\alpha \gets \conjProc{\varphi}{\cL_3}{\cU}$\;
        \If{$\alpha \neq \bot$} {\label{ln:conjpre}
          $k \gets \max \{ j \mid \cO_j \limp \neg \alpha \}$\;
          $\textsc{push}(Q, \langle \alpha, k \rangle)$ \label{ln:conj}\tcp*{Conjecture}
        }
        \If{$\neg \pi_3 = A \cdot \Consts \leq \Vars$} {
$\psi \gets \textsc{subsume}(\langle \pi_3, \cL_3\rangle)$\; \label{ln:sub-call}
          $k \gets \max \{j \mid \cF(\cO_{j}) \limp \psi'\}$\;
          $\cO_j \gets \cO_j \land \psi \textbf{ for all } j \leq k+1$\label{ln:sub}  \tcp*[f]{Subsume}
        }
        \BlankLine
    }
\end{multicols}
\end{algorithm2e}
Having explained the implementation of the new rules for \lia, we now put all
the ingredients together into an algorithm, \csm. In particular, we present our
choices as to when to apply the new rules, and on which clusters of lemmas and
\pobs. As can be seen in \Cref{sec:eval}, this implementation works very
well on a wide range of benchmarks.

\Cref{alg:impl} presents \csm. The comments to the right side of a line refer to
the abstract rules in \Cref{alg:spc,alg:gspacer-rules}. Just like \Spacer, \csm
iteratively computes predecessors~(\cref{ln:pred}) and blocks
them~(\cref{ln:blk}) in an infinite loop. Whenever
a \pob is proven to be reachable, the reachable states are
updated~(line~\ref{ln:succ}). If $\Bad$ intersects with a reachable state, \csm
terminates and returns \unsafe~(line~\ref{ln:unsafe}). If one of the frames is
an inductive invariant, \csm terminates with \safe~(line~\ref{ln:safe}).

When a \pob $\langle \varphi, i \rangle$  is handled, we first apply
the \texttt{Concretize} rule, if possible~(\cref{ln:concrposs}).
Recall that \textsc{Concretize}~(\Cref{alg:gual}) takes as input a cluster that partially blocks $\varphi$
and has a non-linear pattern.
To obtain such a cluster, we first find, using $\pobclstr{\varphi}{i}$, a cluster $\langle \pi_1, \cL_1 \rangle = \cC_{\cO_k}(\pi_1)$, where $k \leq i$,
that includes \emph{some} lemma (from frame $k$)
that blocks $\varphi$; if none exists, $\cL_1 = \emptyset$.
We then filter out from $\cL_1$ lemmas that completely block $\varphi$
as well as lemmas that are irrelevant to $\varphi$, i.e., we obtain $\cL_2$ by keeping only lemmas that
partially block $\varphi$. We apply \textsc{Concretize} on $\langle \pi_1, \cL_2\rangle$
to obtain a new \pob that under-approximates $\varphi$ if \begin{inparaenum}[(1)]
\item the remaining sub-cluster, $\cL_2$, is non-empty,
\item the pattern, $\pi_1$, is non-linear, and
\item  $\Land \cL_2 \land \varphi$ is satisfiable, i.e., a part of $\varphi$ is not blocked by any lemma in $\cL_2$. \end{inparaenum}

Once a \pob is blocked, and a new lemma that blocks it, $\ell$, is added to the
frames, an attempt is made to apply the \texttt{Subsume} and \texttt{Conjecture}
rules on a cluster that includes $\ell$. To that end, the function
$\lclstr{\ell}$ finds \emph{a} cluster $\langle \pi_3, \cL_3 \rangle =
\cC_{\cO_i}(\pi_3)$ to which $\ell$ belongs~(\Cref{sec:clstr}). Note that the choice of cluster is
arbitrary. The rules are applied on $\langle \pi_3, \cL_3 \rangle$ if the
required pre-conditions are met (\cref{ln:conjpre} and \cref{ln:sub-call},
respectively). When applicable, \textsc{Subsume} returns a new lemma that is added to the frames,
while \textsc{Conjecture} returns a new \pob that is added to the queue.
Note that the latter is a \emph{may} \pob, in the sense
that some of the states it represents \emph{may not} lead to safety violation.

\paragraph{Ensuring progress.} \Spacer always makes progress: as its search
continues, it establishes absence of counterexamples of deeper and deeper
depths. However, \csm does not ensure progress. Specifically, unrestricted
application of the \texttt{Concretize} and \texttt{Conjecture} rules can make
\csm diverge even on executions of a fixed bound. In our implementation, we
ensure progress by allotting a fixed amount of \emph{gas} to each pattern,
$\pi$, that forms a cluster. Each time \texttt{Concretize} or
\texttt{Conjecture} is applied to a cluster with $\pi$ as the pattern, $\pi$
loses some gas. Whenever $\pi$ runs out of gas, the rules are no longer applied
to any cluster with $\pi$ as the pattern. There are finitely many patterns
(assuming LIA terms are normalized). Thus, in each bounded execution of \csm,
the \texttt{Concretize} and \texttt{Conjecture} rules are applied only a finite
number of times, thereby, ensuring progress. Since the \texttt{Subsume} rule
does not hinder progress, it is applied without any restriction on gas.

  \section{Evaluation}
\setlength{\textfloatsep}{20pt}
\label{sec:eval}
We have implemented\footnote{\url{https://github.com/hgvk94/z3/tree/gspacer-cav-ae}} \GSpacer~(\Cref{alg:impl}) as an extension to \Spacer.
To reduce the dimension of a matrix (in \textsc{subsume}, \Cref{sec:mrg}), we
compute pairwise linear dependencies between all pairs of columns instead of
computing the full kernel. This does not necessarily reduce the dimension of the
matrix to its rank, but, is sufficient for our benchmarks. We have experimented
with computing the full kernel using SageMath~\cite{sagemath}, but the overall
performance did not improve. Clustering is implemented by anti-unification. LIA
terms are normalized using default Z3 simplifications. Our implementation also
supports global generalization for non-linear CHCs. We have also extended our
work to the theory of LRA. We defer the
details of this extension to an extended version of the
paper.

To evaluate our implementation, we have conducted two sets of
experiments\footnote{\label{web}Detailed experimental results including the
  effectiveness of each rule, and the extensions to non-linear CHCs and LRA can
  be found at \url{https://hgvk94.github.io/gspacer/}}. All experiments were run
on Intel E$5$-$2690$ V$2$ CPU at $3$GHz with $128$GB memory with a timeout of
$10$ minutes. First, to evaluate the performance of local reasoning with global
guidance against pure local reasoning, we have compared \GSpacer with the latest
\Spacer, to which we refer as the \emph{baseline}. We took the benchmarks from
CHC-COMP 2018 and~2019~\cite{chc-comp}. We compare to \Spacer because it dominated the competition by solving
$85\%$ of the benchmarks in CHC-COMP 2019 ($20\%$ more than the
runner up) and $60\%$ of the benchmarks in CHC-COMP 2018 ($10\%$ more than
runner up). Our evaluation shows that \GSpacer outperforms \Spacer both in terms
of number of solved instances and, more importantly, in overall robustness.

Second, to examine the performance of local reasoning with global guidance
compared to solely global reasoning,
we have compared \GSpacer with an ML-based
data-driven invariant inference tool \textsc{LinearArbitrary}~\cite{DBLP:conf/pldi/ZhuMJ18}.
Compared to other similar
approaches, \textsc{LinearArbitrary} stands out by supporting invariants with arbitrary Boolean
structure over arbitrary linear predicates. It is completely automated and does
not require user-provided predicates, grammars, or any other guidance.
For the comparison with \textsc{LinearArbitrary}, we have used both the CHC-COMP
benchmarks, as well as the benchmarks from the artifact evaluation
of~\cite{DBLP:conf/pldi/ZhuMJ18}. The machine and timeout remain the same.
Our evaluation shows that \GSpacer is superior in this case as well.

\paragraph{Comparison with \Spacer.}
\Cref{tbl:csmvspc} summarizes the comparison between \Spacer and \GSpacer on
CHC-COMP instances. Since both tools can use a variety of interpolation
strategies during lemma generalization~(\Cref{ln:indgen} in~\Cref{alg:impl}), we
compare three different configurations of each: \emph{bw} and \emph{fw} stand
for two interpolation strategies, \emph{backward} and \emph{forward},
respectively, already implemented in \Spacer, and \emph{sc} stands for turning
interpolation off and generalizing lemmas only by \emph{subset clauses} computed
by inductive generalization.

\begin{figure}[t]
  \centering
  \subcaptionbox{\label{fig:csmvsspctime} running time}{
      \includegraphics[width = 0.45\textwidth]{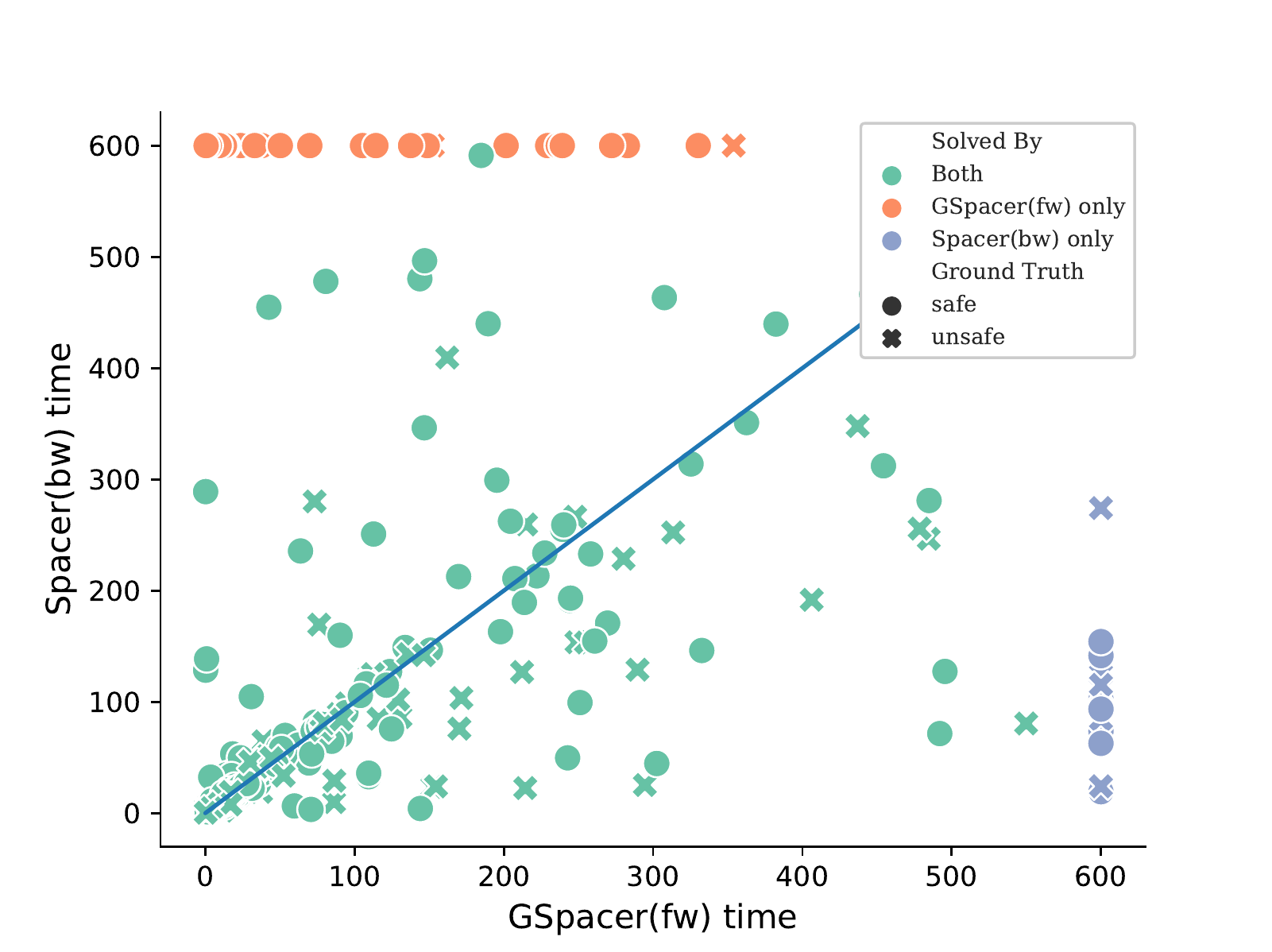}}
    \hfill
\subcaptionbox{\label{fig:csmvsspcdepth} depth explored}{
    \includegraphics[width = 0.45\textwidth]{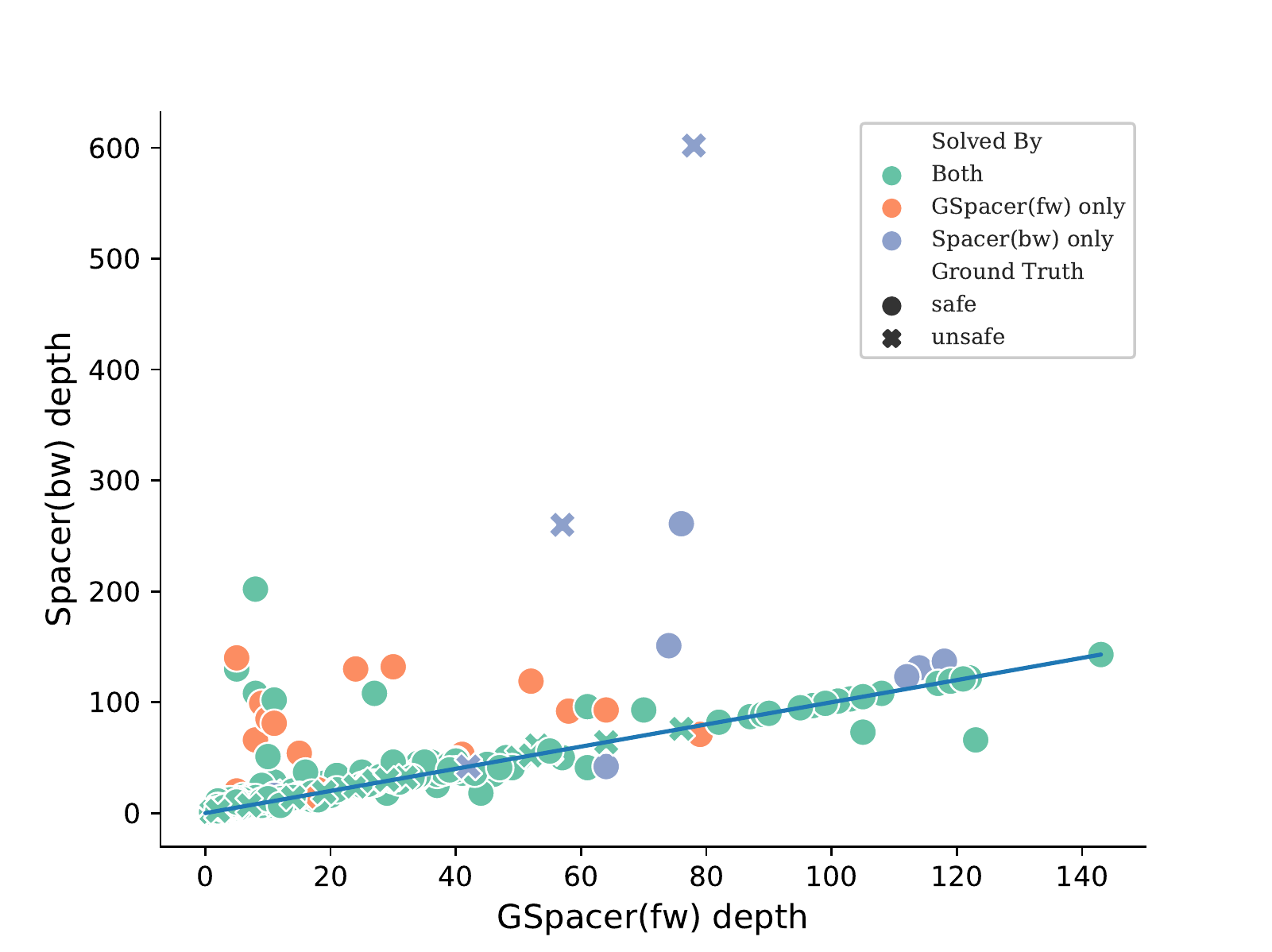}}
  \vspace{-0.1in}
  \caption{Best configurations: \GSpacer versus \Spacer.}
  \label{fig:csmvspc}
  \vspace{-0.3in}
\end{figure}

Any configuration of \GSpacer solves significantly more instances than
even the best configuration of \Spacer. \Cref{fig:csmvspc} provides a more
detailed comparison between the best configurations of both tools in terms of
running time and depth of convergence. There is no clear trend in terms of
running time on instances solved by both tools. This is not surprising ---
SMT-solving run time is highly non-deterministic and any change in strategy has
a significant impact on performance of SMT queries involved. In terms of depth,
it is clear that \GSpacer converges at the same or lower depth. The depth is
significantly lower for instances solved only by \GSpacer.

Moreover, the performance of \GSpacer is not significantly affected by
the interpolation strategy used. In fact, the configuration \emph{sc} in which
interpolation is disabled performs the best in CHC-COMP 2018, and only slightly
worse in CHC-COMP 2019! In comparison, disabling interpolation hurts \Spacer significantly.

\begin{figure}[t]
  \centering
  \subcaptionbox{\label{fig:csmvcsmfwtime} running time}{
    \includegraphics[width = 0.4\textwidth]{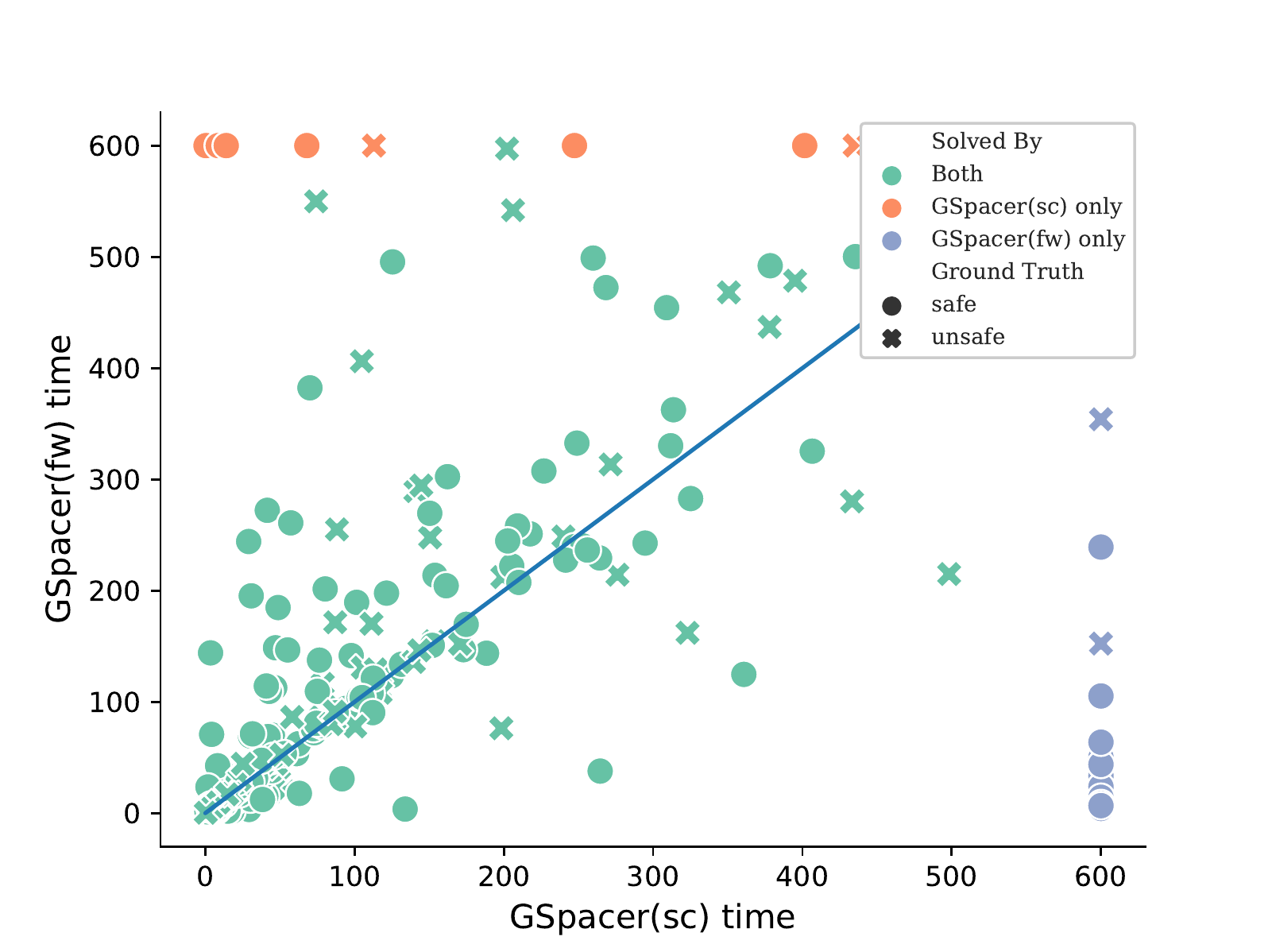}}
  \subcaptionbox{\label{fig:csmvcsmfwdepth} depth explored}{
    \includegraphics[width = 0.4\textwidth]{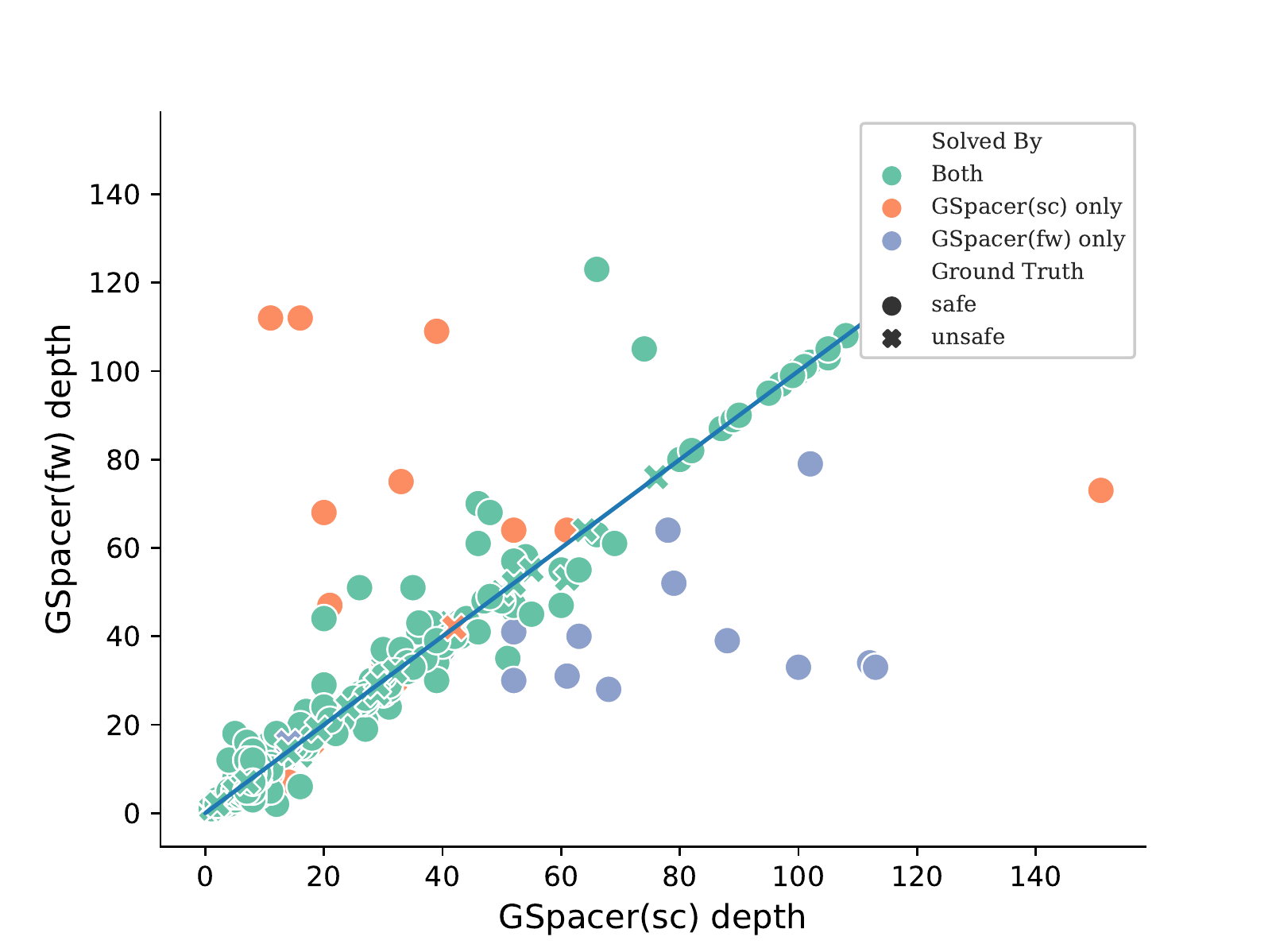}}\\
    \caption{Comparing \GSpacer with different interpolation tactics.}
\label{fig:csmvcsm}
\end{figure}

\Cref{fig:csmvcsm} provides a detailed comparison of \GSpacer with and without
interpolation. Interpolation makes no difference to the depth of convergence.
This implies that lemmas that are discovered by interpolation are discovered as
efficiently by the global rules of \GSpacer. On the other hand, interpolation
significantly increases the running time. Interestingly, the time spent in interpolation itself is
insignificant. However, the lemmas produced by interpolation tend to slow down
other aspects of the algorithm. Most of the slow down is in increased time for
inductive generalization and in computation of predecessors. The comparison
between the other interpolation-enabled strategy and \mbox{\GSpacer(\emph{sc})}
shows a similar trend.

\definecolor{Gray}{gray}{0.85}
\definecolor{LightCyan}{rgb}{0.88,1,1}
\newcolumntype{a}{>{\columncolor{Gray}}c}

\begin{table}[t]
  \centering
  \small
  \begin{adjustbox}{max width=1.2\textwidth}
    \begin{tabular}{@{}cacacac||acacac||ac@{}}
      \toprule

      \mc{1}{c}{ \multirow{2}{*}{\textbf{Bench}}  }& \mc{6}{c}{\Spacer} & \mc{6}{c}{\GSpacer}\\
      \cmidrule(lr){2-7}
      \cmidrule(lr){8-13}
       & \mc{2}{c}{fw} & \mc{2}{c}{bw} & \mc{2}{c}{sc} & \mc{2}{c}{fw} & \mc{2}{c}{bw} & \mc{2}{c}{sc} & \mc{2}{c}{VBS} \\
      \midrule

      \rowcolor{White}
      & safe & unsafe & safe & unsafe & safe & unsafe & safe & unsafe & safe & unsafe & safe & unsafe & safe & unsafe \\
      \textbf{CHC-18} & 159 & 66 & 163 & \textbf{69} & 123 & 68 & \textbf{214} & 67 & \textbf{214} & 63 & \textbf{214} & \textbf{69} & 229 & 74 \\

      \textbf{CHC-19} & 193 & 84 & 186 & 84 & 125 & 84 & \textbf{202} & 84 & 196 & \textbf{85} & 200 & 84 & 207 & 85\\

      \bottomrule
    \end{tabular}
    \end{adjustbox}
    \vspace{0.1in}
  \caption{Comparison between \Spacer and \GSpacer on CHC-COMP.}
  \label{tbl:csmvspc}
  \vspace{-0.3in}
\end{table}

\paragraph{Comparison with \textsc{LinearArbitrary}.} In~\cite{DBLP:conf/pldi/ZhuMJ18}, the authors show that \textsc{LinearArbitrary},
to which we refer as \ddchc for short, significantly
outperforms \Spacer on a curated subset of benchmarks from SV-COMP~\cite{svcomp}
competition. 

At first, we attempted to compare \ddchc against \GSpacer on the CHC-COMP
benchmarks. However, \ddchc did not perform well on them. Even the baseline
\Spacer has outperformed \ddchc significantly.
Therefore, for a more meaningful comparison, we have also compared \Spacer, \ddchc and
\GSpacer on the benchmarks from the artifact evaluation
of~\cite{DBLP:conf/pldi/ZhuMJ18}. The results are summarized in
\Cref{tbl:ddchc}. As expected, \ddchc outperforms the baseline \Spacer on the
safe benchmarks. On unsafe benchmarks, \Spacer is significantly better than \ddchc. In both
categories, \GSpacer dominates solving more safe benchmarks than either \Spacer
or \ddchc, while matching performance of \Spacer on unsafe instances.
Furthermore, \GSpacer remains orders of magnitude faster than \ddchc on
benchmarks that are solved by both. \ag{missing: gspacer unique 17, ddhch unique
  5}
This comparison shows that incorporating local reasoning with global guidance
  not only mitigates its shortcomings but also surpasses global data-driven reasoning.

\begin{table}[t]
  \centering
  \small
  \begin{adjustbox}{max width=1.2\textwidth}
\begin{tabular}{@{}cacacacac@{}}
\toprule
  \textbf{Bench} & \mc{2}{c}{\Spacer} & \mc{2}{c}{\ddchc} & \mc{2}{c}{\GSpacer} & \mc{2}{c}{VB}\\
  \hline

  & safe & unsafe & safe & unsafe & safe & unsafe & safe & unsafe\\

  \textbf{PLDI18} & 216 & \textbf{68} & 270 & 65  & \textbf{279} & \textbf{68}  & 284 & 68\\
  \hline\\
\end{tabular}
\end{adjustbox}
\caption{Comparison with \ddchc.}
\label{tbl:ddchc}
\vspace{-0.3in}
\end{table}

\vspace{-15pt}
 \section{Related Work}
\label{sec:related}

The limitations of local reasoning in SMT-based infinite state model checking are
well known. Most commonly, they are addressed with either (a)~different strategies
for local generalization in interpolation (e.g.,
\cite{DBLP:journals/acta/LerouxRS16,DBLP:conf/cav/AlbarghouthiM13,DBLP:conf/tacas/BlichaHKS19,DBLP:conf/vmcai/SchindlerJ18}),
or (b)~shifting the focus to \emph{global} invariant inference by learning an
invariant of a restricted shape~(e.g.,
\cite{DBLP:conf/fm/FlanaganL01,DBLP:conf/pldi/ZhuMJ18,DBLP:conf/popl/0001NMR16,DBLP:conf/fmcad/FedyukovichKB17,DBLP:conf/tacas/ChampionC0S18}).

\paragraph{Interpolation strategies.}
Albarghouthi and McMillan~\cite{DBLP:conf/cav/AlbarghouthiM13} suggest to
minimize the number of literals in an interpolant, arguing that simpler (i.e.,
fewer half-spaces) interpolants are more likely to generalize. This helps with
myopic generalizations (\Cref{ex:mrg}), but not with excessive generalizations
(\Cref{ex:splt}). On the contrary, Blicha et
al.~\cite{DBLP:conf/tacas/BlichaHKS19} decompose interpolants to be numerically
simpler (but with more literals), which helps with excessive, but not with
myopic, generalizations. Deciding \emph{locally} between these two techniques or on
their combination (i.e., some parts of an interpolant might need to
be split while others combined) seems impossible. Schindler and
Jovanovic~\cite{DBLP:conf/vmcai/SchindlerJ18} propose local interpolation that
bounds the number of lemmas generated from a single \pob (which helps with
\Cref{ex:conjecture}), but only if inductive generalization is disabled.
Finally,~\cite{DBLP:journals/acta/LerouxRS16} suggests using external guidance,
in a form of predicates or terms, to guide interpolation. In contrast, \GSpacer
uses global guidance, based on the current proof, to direct different local
generalization strategies. Thus, the guidance is automatically tuned to the
specific instance at hand rather than to a domain of problems.

\paragraph{Global invariant inference.} An alternative to inferring lemmas for
the inductive invariant by blocking counterexamples is to enumerate the space of
potential candidate
invariants~\cite{DBLP:conf/fm/FlanaganL01,DBLP:conf/pldi/ZhuMJ18,DBLP:conf/popl/0001NMR16,DBLP:conf/fmcad/FedyukovichKB17,DBLP:conf/tacas/ChampionC0S18}.
This does not suffer from the pitfall of local reasoning. However, it is only
effective when the search space is constrained. While these approaches
perform well on their target domain, they do not generalize well to a diverse
set of benchmarks, as illustrated by results of CHC-COMP and our empirical
evaluation in~\Cref{sec:eval}.

\paragraph{Locality in SMT and \textsc{IMC}.} Local reasoning is also a known
issue in SMT, and, in particular, in DPLL(T)
(e.g.,~\cite{DBLP:conf/cav/McMillanKS09}). However, we are not aware of global
guidance techniques for SMT solvers. Interpolation-based Model Checking
(IMC)~\cite{DBLP:conf/cav/McMillan03,DBLP:conf/cav/McMillan06} that uses
interpolants from proofs, inherits the problem. Compared to IMC, the propagation
phase and inductive generalization of \Ict~\cite{DBLP:conf/vmcai/Bradley11}, can
be seen as providing global guidance using lemmas found in other parts of the
search-space. In contrast, \GSpacer magnifies such global guidance by exploiting
patterns within the lemmas themselves.

\paragraph{IC3-SMT-based Model Checkers.}
There are a number of IC3-style SMT-based infinite state model checkers,
including~\cite{DBLP:conf/cav/KomuravelliGC14,DBLP:conf/fmcad/JovanovicD16,DBLP:journals/fmsd/CimattiGMT16}.
To our knowledge, none extend the IC3-SMT framework with a global guidance. A
rule similar to \texttt{Subsume} is suggested in~\cite{DBLP:conf/date/WelpK13}
for the theory of bit-vectors and in~\cite{DBLP:conf/vmcai/BjornerG15} for LRA,
but in both cases without global guidance. In~\cite{DBLP:conf/vmcai/BjornerG15},
it is implemented via a combination of syntactic closure with interpolation,
whereas we use MBP instead of interpolation. Refinement State Mining
in~\cite{DBLP:conf/cav/BirgmeierBW14} uses similar insights to our  \texttt{Subsume} rule to refine predicate abstraction. \vspace{-15pt}

 \section{Conclusion and Future Work}

This paper introduces \emph{global guidance} to mitigate the limitations of the
local reasoning performed by \smt-based \Ict-style model checking algorithms.
Global guidance is necessary to redirect such algorithms from divergence due to
persistent local reasoning. To this end, we present three general rules that
introduce new lemmas and \pobs by taking a global view of the lemmas learned so
far. The new rules are not theory-specific, and, as demonstrated by
\Cref{alg:impl}, can be incorporated to \Ict-style solvers without modifying
existing architecture. We instantiate, and implement, the rules for \lia in
\GSpacer, which extends \Spacer.

Our evaluation shows that global guidance brings significant improvements to
local reasoning, and surpasses invariant inference based solely on global reasoning.
More importantly, global guidance decouples \Spacer's dependency on
interpolation strategy and performs almost equally well under all three
interpolation schemes we consider. As such, using global guidance in the context
of theories for which no good interpolation procedure exists, with bit-vectors
being a primary example, arises as a promising direction for future research.

\vspace{-15pt}
\label{sec:concl}

 \subsubsection*{Acknowledgements}
We thank Xujie Si for running the \ddchc experiments and
collecting results.
We thank the ERC starting Grant SYMCAR 639270 and the Wallenberg Academy
Fellowship TheProSE for supporting the research visit.
This research was partially supported by the United States-Israel Binational Science Foundation (BSF) grant No.\ 2016260,
and the Israeli Science Foundation (ISF) grant No.\ 1810/18. This research was
partially supported by grants from Natural Sciences and Engineering Research Council
Canada.

 \bibliographystyle{abbrv}
\bibliography{ref}
\end{document}